\documentclass{aa}
\usepackage{graphicx}
\usepackage{caption}
\usepackage[varg]{txfonts}
\begin{document}
\title{A parameterisation of the longitudinal Cherenkov emission profiles of gamma induced electromagnetic showers in the atmosphere in the GeV-TeV energy range}

\author{Saeeda Sajjad 
\inst{1}
\inst{2}
\and 
Alain Falvard
\inst{2}}

\institute{Space and Astrophysics Research Lab (SARL), National Centre of GIS and Space Applications (NCGSA), Institute of Space Technology, Islamabad 44000, Pakistan\\
\email{saeeda.sajjad@mail.ist.edu.pk}
\and Laboratoire  Univers  et  Particules  de  Montpellier (LUPM) CNRS/Universit\'{e}  de  Montpellier (UMR-5299),  Place  E.~Bataillon, 34095 Montpellier, France}

\abstract{In gamma-ray astronomy through Imaging Atmospheric Cherenkov Telescopes (IACT), the atmosphere is used as a calorimeter. Incident gamma-rays in the GeV-TeV energy range are observed through the electromagnetic showers they produce in the atmosphere and the Cherenkov light emitted by them. 
}
{We aim to obtain a parameterisation of the longitudinal profiles of electromagnetic showers in the atmosphere. The longitudinal Cherenkov emission profiles of these showers can be used for the reconstruction of the parameters of the gamma-photon as well gamma-hadron discrimination. Obtaining an accurate parameterisation of these profiles can therefore be used in the analysis of IACT images as well as fast simulation of the Cherenkov emission from gamma-induced showers in the atmosphere. Such parameterisations are not know to have been systematically carried out prior to this.}
{We carry out Monte Carlo simulations of gamma-induced showers in the atmosphere. The parameterisation of their Cherenkov emission longitudinal profiles is based on the well-known gamma-function used in particle detector physics. }
{We evaluate the dependence of the depth of the maximum and the average depth of the longitudinal Cherenkov emission profile, as well as the parameter $1/\beta$ on the energy of the initial shower. We also study the distributions of various parameters associated with the gamma-function and find that $\beta/\alpha$ and $1/\alpha$ have Gaussian distributions and are not correlated with each other which makes them appropriate variables for evaluating the fluctuations of the Cherenkov emission profiles. We finally present parameterisations of the fluctuations of these two variables.}
{}

%\keywords{imaging atmospheric Cherenkov telescopes - IACT - Cherenkov - longitudinal profile - gamma-ray astronomy}
\keywords{gamma rays: general - methods: numerical}

\titlerunning{Longitudinal profile parameterisation of EM showers in the atmosphere}
\authorrunning{S. Sajjad et al.}

\maketitle 

\section{Introduction}

In Very High Energy (VHE) astronomy, from a few tens of GeVs to a few tens of TeVs, observations of gamma-rays are carried out predominantly through Imaging Atmospheric Cherenkov Telescopes (IACT). Currently, three major observatories HESS, MAGIC and VERITAS are carrying out gamma-ray observations in this energy range \citep{aharonian2006observations, magic2005,weekes2002veritas}. The future observatory CTA is also being developed \citep{actis2011design} and is expected to increase the number of observable VHE sources in the sky by ten folds.

The IACT technique consists of using the atmosphere as a calorimeter. In the VHE range, the gamma-ray flux from astrophysical sources is very low. Since space-based observatories have limited surface areas, the possibility of observing such sources in space is severely limited. VHE gamma-rays are instead observed on the ground through the IACT technique. The gamma-ray produces an electromagnetic shower in the atmosphere and the Cherenkov light emitted by the charged particles of this shower is used to obtain images of the shower. The analysis of these images allows for the reconstruction of the original gamma-ray as well as the rejection of hadronic background.

Given the importance of IACT arrays in developing our understanding of high energy phenomena in the Universe IACT analysis and simulation methods undergo regular improvement and diversification.

In this study, we focus on the longitudinal Cherenkov emission profiles of electromagnetic showers in the atmosphere. These profiles have the potential to be used for shower reconstruction, gamma-hadron discrimination as well as fast simulations. The well-known gamma-function by \cite{longo1975monte} is frequently used to represent the longitudinal energy deposition profiles of electromagnetic showers in calorimeters of various materials in particle detectors. However, a systematic parameterisation of the longitudinal Cherenkov emission profiles of gamma-ray showers in the atmosphere through the gamma function is not known to have been carried out up to this point.

Using an approach similar to the one used by \cite{grindhammer2000parameterized} and \cite{radel2013calculation}, we carry out the parameterisations of the longitudinal profiles of simulated electromagnetic showers in the atmosphere through the gamma function. However, contrary to them, we focus on the longitudinal profiles of the Cherenkov emission from the simulated gamma-induced showers since they are relevant to observations with IACT. The longitudinal Cherenkov emission profiles are expected to also be well represented by the gamma-function since Cherenkov light is emitted by the charged particles in the shower whenever their speed is greater than the speed of light in the air. The distribution of the charged particles itself determines the energy deposition profile in the atmosphere.

The parameterisation of longitudinal Cherenkov emission profiles is of interest for IACT since these profiles can be reconstructed from IACT images. Longitudinal profiles of electromagnetic showers are sensitive to various parameters of the primary gamma-ray such as its energy, depth of first interaction, angle of inclination. Moreover, the longitudinal profiles of gamma and hadron showers are also different from each other and can provide a way to carry out gamma-hadron discrimination. The parameterisation obtained in this study can therefore be used to analyse the data from IACT. They can also be used in fast simulations for IACT arrays.

This paper is organised in the following way. Section \ref{sec_longiprofiles_modelling} discusses the use of longitudinal profiles of atmospheric showers in astrophysics, followed by a discussion on the modelling of electromagnetic showers. We also give a description of the gamma-function and parameters associated with it. Section \ref{sec_showers_atmosphere} gives a brief overview of electromagnetic shower development in the atmosphere, with section \ref{subsec_atmosphere} focusing on the particularities of using the atmosphere as a calorimeter. This is followed by section \ref{subsec_Cherenkov} which describes briefly the Cherenkov emission in electromagnetic showers in the atmosphere.

In section \ref{sec_simulations}, we describe the Monte Carlo simulations through CORSIKA carried out for this study. In particular, we study the effect of using a minimum energy threshold for tracking particles in these simulations in section \ref{sec_ecuts} and discuss potential effects on the results. We also describe how the number of showers simulated at each energy was chosen as a function of the fluctuations in the showers.

Finally, section \ref{sec_results} gives the results of the parameterisations. We also give an evaluation of the fluctuations of various parameters and the correlations between them.  Section \ref{sec_conclusions} gives a summary of the results as well as conclusions and future directions.

\section{Longitudinal profiles and modelling electromagnetic showers}\label{sec_longiprofiles_modelling}

\subsection{The use of longitudinal profiles of atmospheric showers in astrophysics}

 The use of the longitudinal information of particle showers is of importance in several areas of research. For instance, in the case of observations of Ultra High Energy Cosmic Rays (UHECR), they have been studied extensively to develop a better understanding of the processes involved in these showers as well as to reconstruct information about the primary particle. Longitudinal profiles and positions of shower maximum of UHECR showers have been measured by \cite{abu2001measurement}, \cite{guy2002comparison}, \cite{abraham2010measurement}, \cite{kampert2012measurements}, \cite{aab2014depth}, \cite{aab2019measurement} and \cite{andringa2019average}. The profiles have also been studied to improve reconstruction and analysis methods (see for instance \cite{song2004longitudinal}, \cite{unger2008reconstruction}, \cite{andringa2011mass}, \cite{baus2011anomalous},\cite{conceiccao2015average}, \cite{muller2019longitudinal}).

In IACT arrays, the information of the development of the profiles is usually accessed indirectly through Hillas parameters characterising a Cherenkov image \citep{hillas1985cerenkov}. Another analysis method, known as the 3D-Model analysis for IACT by \cite{lemoine2006selection}, models the electromagnetic shower as a 3-dimensional photosphere. The longitudinal (as well as lateral) profile of this photosphere is modelled as a Gaussian. \cite{TheseMathieudeNaurois} and \cite{de2015ground} have shown a logarithmic dependence on the energy for the depth of the maximum of shower development. In the former, simulated gamma-ray showers in the atmosphere are fitted whereas in the latter, an analytical solution based on the Greisen function is used.

\subsection{Modelling electromagnetic shower development}

In comparison with hadronic showers (and in particular showers of ultra high energy cosmic rays), the development of electromagnetic showers is relatively simple and the processes involved are quite well understood. Models to describe the development of electromagnetic showers have been evolving  since the 1930s.

The development of electromagnetic showers has been described analytically through various approaches. Important works of note include the calculations by \cite{bhabha1937passage}, \cite{carlson1937multiplicative}, the use of diffusion equations by  \cite{rossi1941cosmic} and \cite{Kamata1958Lateral}, and the development of functions by \cite{greisen1956progress} and \cite{gaisser1977reliability}. More simplified and intermediate approaches include the ones by \cite{heitler1954quantum}, \cite{rossi1952}, and \cite{montanus2012intermediate}.

Empirically, the gamma function derived by \cite{longo1975monte} is widely used to describe the longitudinal profiles of electromagnetic showers in particle detectors. The average longitudinal profile for energy deposition by an electromagnetic shower in the medium is known to be well represented by this function. Simulated longitudinal profiles of electromagnetic showers have been fitted and parameterised by the gamma-function (see for instance \cite{PhysRevD.98.030001}). The gamma function has also been used by \cite{grindhammer2000parameterized} to obtain parameterisations of electromagnetic showers in various materials in the case of homogeneous and non-homogeneous materials. Their work also includes the study of the fluctuations of the showers and study the correlations between various parameters. Following \cite{grindhammer2000parameterized}, \cite{radel2013calculation} use Geant4 simulated electromagnetic showers to obtain parameterisations of their longitudinal profiles for various energies in ice and water for neutrino observatories such as IceCube and Antares \citep{aartsen2017icecube,ageron2011antares}. The gamma function has also been used by for fast simulations of electromagnetic showers for particle detectors (See for example \cite{grindhammer1989fast}, \cite{Jang2009}).

\subsection{The gamma function}\label{subsec_gamma}

The gamma function gives the energy deposited as a function of the depth in the atmosphere through the expression:
\begin{equation}\label{eq_gamma}
\frac{dE}{dt}=E_0 \beta \frac{(\beta t)^{\alpha-1}e^{-\beta t}}{\Gamma(\alpha)}.
\end{equation}

Here, $\Gamma(\alpha)=\int_0^{\infty} e^{-z}z^{\alpha-1}dz$ and $E_0$ is the energy of the primary particle. In addition, the depth $t$ in the atmosphere is expressed in units of radiation lengths, such that $t=z/X_0$, where $X_0$ is the radiation length (equal to 36.66~g/cm$^2$ for air) and $z$ is the mass thickness expressed in g/cm$^2$. Expressing the depth in terms of mass thickness and units of radiation lengths makes the expression independent of the atmospheric model and composition. When $t$ is small, the factor $(\beta t)^{\alpha-1}$ dominates the profile of energy deposition. At this stage, the number of charged particles in the shower is limited to a few. The energy deposition occurs through the loss of energy of these charged particles through bremsstrahlung. The parameter $\alpha$, therefore, governs the initial part of the shower development. After a few radiation lengths, the energy deposit increases as the number of particles in the showers multiplies. At this point, the profile is no longer determined by $\alpha$ alone and the value of $\beta$ also plays a role in it. When $t$ is very large, the factor $e^{-\beta t}$ determines the energy deposition profile. At this point, the number of particles in the shower is decaying due to the absorption in the atmosphere. This part of the profile is described by the parameter $\beta$.

Using the gamma function, the depth of the shower maximum, the shower's centre of gravity and variance can be calculated as
\begin{eqnarray}
	&t_{max}&=\frac{\alpha-1}{\beta},\label{eq_tmax}\\ 
	&\langle t\rangle&=\frac{\alpha}{\beta},\label{eq_tmean}\\ 
	&\langle t^2\rangle&=\frac{\alpha(\alpha-\beta)}{\beta^2},
\end{eqnarray}
respectively. One can also show that the higher order moments are given by 
\begin{eqnarray*}
	\langle t^n\rangle=\frac{(\alpha+n-1)\cdots\alpha}{\beta^n}.
\end{eqnarray*}

In order to further remove the dependence on the atmospheric profile, the energy $E$ is expressed in units of the critical energy $E_c$, through $y=E/E_c$. The critical energy is around 86~MeV for air. The use of these units for length and energy is also relevant to the 'Approximation B' of Rossi's model \citep{rossi1952} which predicts that electromagnetic shower profiles are independent of material if such units are used.

\begin{figure*}
\begin{center}
\includegraphics[width=0.85\textwidth,keepaspectratio]{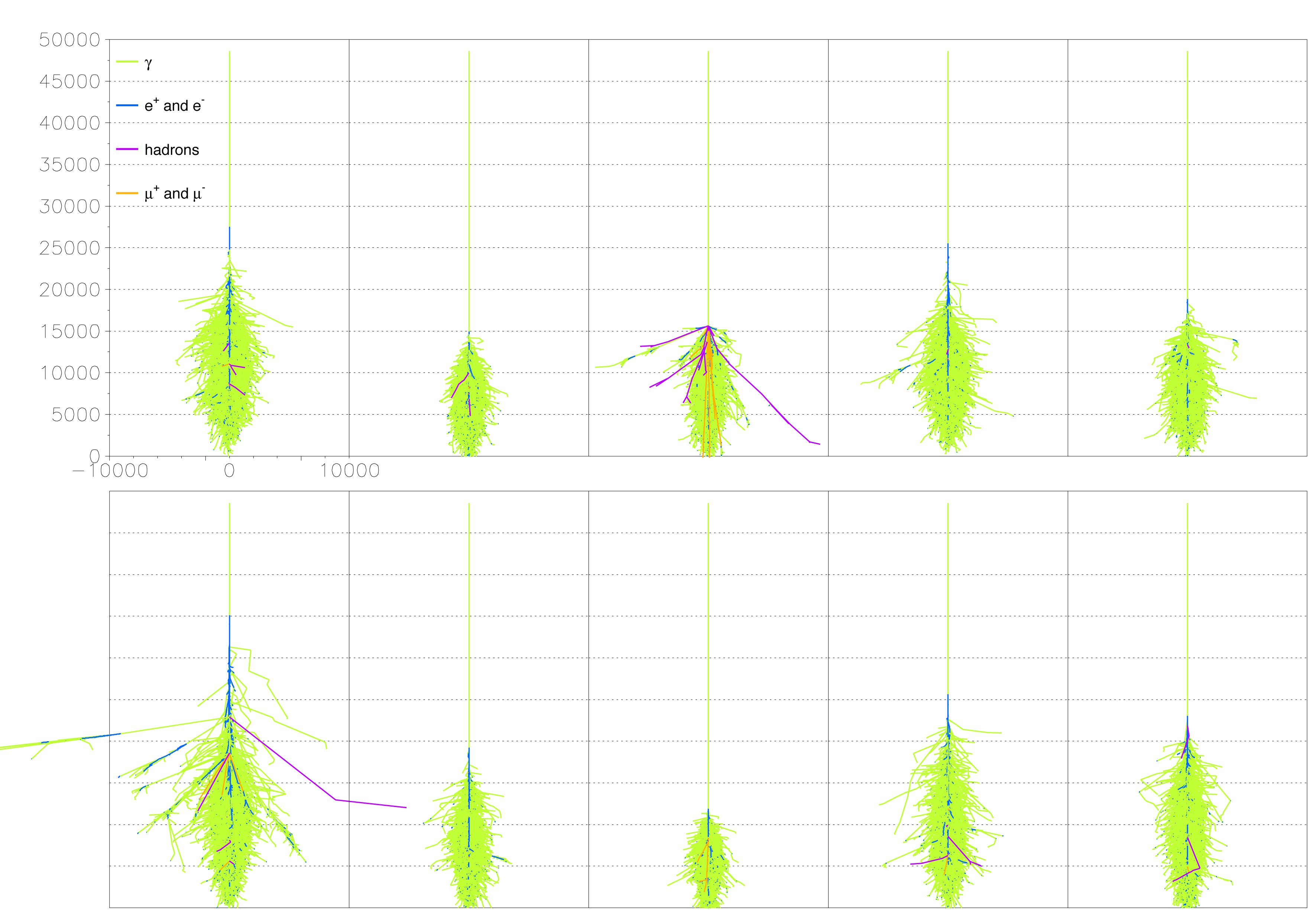}
\caption{\label{fig_show_phE500_10show}Ten simulated gamma-initiated showers of 500~GeV. The units of the vertical and horizontal axes are metres. The horizontal scale is deliberately chosen to be larger so as to emphasise the longitudinal development of the shower. Figure from \cite{sajjad:tel-00408835}.}
\end{center}
\end{figure*}
\section{Electromagnetic showers in the atmosphere and their Cherenkov emission} \label{sec_showers_atmosphere}

When a VHE gamma-ray enters the atmosphere, its predominant mode of interaction with the surrounding matter is through pair production. The electron and positron resulting from this process undergo bremsstrahlung as they interact with the particles of the atmosphere. The repetition of these two processes results in the exponential growth of the number of particles and the production of an electromagnetic shower in the atmosphere. This phase of shower development corresponds to the high energy regime. During this phase, as the number of particles grows, the energy of individual particles decreases. When the energy of charged particles lowers to values below the critical energy of air, the predominant mode of interaction of these particles with the surrounding material changes from bremsstrahlung to ionisation. As a result, the particles of the shower start getting absorbed and the total number of particles in the shower starts diminishing. This marks the beginning of the low energy regime of the shower. The shower maximum occurs prior to this, when the number of particles in the shower is the highest.

Apart from pair production and bremsstrahlung, other processes that play a role in the development of include multiple scattering by the charged particles, Compton effect at lower photon energies and occasional muon pair production and photoproduction. 
In addition, the magnetic field of the Earth can also deflect the charged particles in the shower and therefore have an impact on shower morphology.

Figure \ref{fig_show_phE500_10show} shows ten simulated gamma-ray showers in the atmosphere. The shower development shows fluctuations, specially as a function of the height of first interaction in the atmosphere. This illustrates the need for evaluating fluctuations when carrying out parameterisations for the longitudinal profiles of showers.

\subsection{The atmosphere as a calorimeter for IACT}\label{subsec_atmosphere}

In comparison with particle detector calorimeters whose composition and structure are completely known and optimised according to detection needs, the atmosphere is a complex system.

 Atmospheric density varies greatly with altitude and has an important impact on the longitudinal development of the shower. Even in a simplified isothermal model of the atmosphere, the density decreases exponentially with the altitude. In more complex models, the atmosphere is divided into different layers based on the variation of temperature with altitude. Then, the density profile also follows different parameterisations in each layer. In addition to this, the thickness and layers of the atmosphere are both time and location dependent. Seasons and weather changes as well as geographical location on the Earth have an impact on the profile of the atmosphere.

The composition of the atmosphere also varies with the layers. While its main component is air,  it also contains other elements such as water in the form of clouds and air moisture, aerosols, ozone, etc. Water and aerosols are mainly present in the troposphere while ozone occurs in the stratosphere.

All these factors and variations make the atmosphere a complex, inhomogeneous calorimeter. Observations of gamma-rays with IACT can be significantly affected by factors such as the presence of dust, aerosols and water vapours. IACT observatories continuously monitor the atmosphere in order to account for these factors. A detailed discussion on the subject can be found in \cite{doro2015strategy}, for example.

\subsection{Cherenkov light production in electromagnetic showers in the atmosphere and  gamma-ray observations} \label{subsec_Cherenkov}
The charged particles of the electromagnetic showers produced by gamma-rays have speeds faster than the speed of light in the air. This results in the emission of Cherenkov light from the showers. IACT telescopes collect this Cherenkov light to obtain images of the shower. These images are subsequently analysed to reconstruct the initial gamma-ray and to carry out gamma-hadron separation.

 For a particle of charge $ze$ and speed $\beta$ (in units of the speed of light in vacuum), the number of Cherenkov photons emitted per unit path length $dx$ per unit wavelength interval $d\lambda$ is given by

\begin{equation}\label{eq_ncher_per_length}
\frac{d^2N}{dxd\lambda}=\frac{2\pi\alpha z^2}{\lambda^2}(1-\frac{1}{\beta^2\eta^2})=\frac{2\pi\alpha z^2}{\lambda^2}\mathrm{sin}^2\,\theta_c.
\end{equation}
Here $\alpha$ is the fine structure constant, $\lambda$ the wavelength of the emitted Cherenkov photons, and $\eta$ the refractive index of the atmosphere. $\theta_c$ is the angle of emission of the Cherenkov photon with respect to the path of the charged particle and it is given by $\mathrm{cos}\,\theta_c=\frac{1}{\beta\eta}$.  The threshold energy of Cherenkov emission depends on the mass $m_0$ of the charged particle and the refractive index of air through
\begin{equation}\label{eq_cher_ethr}
E_{thr}=\frac{m_0 c^2}{\sqrt{1-\frac{1}{\eta^2}}}.
\end{equation}
The angle of Cherenkov light emission, the number of Cherenkov photons emitted as well as the energy threshold of emission, all depend on the refractive index of the atmosphere. Since the refractive index of the atmosphere varies with altitude, the Cherenkov emission from the electromagnetic shower is also altitude dependent. The collection of Cherenkov light by IACT is also highly sensitive to atmospheric conditions as water vapours and aerosols can scatter or absorb the Cherenkov photons emitted by the shower.

\section{Gamma-shower simulations}\label{sec_simulations}
We carried out Monte Carlo simulations of electromagnetic showers produced by gamma-rays in the atmosphere through the package CORSIKA, version 6.020 \citep{corsika_physics}. CORSIKA is accepted as the standard tool for the simulation of air showers in astroparticle physics. It uses the package EGS4 to carry out simulations of electromagnetic interactions \citep{egs4}. In addition, CORSIKA also takes into account minor processes involving muons in electromagnetic showers. 

The emission and propagation of Cherenkov light in the atmosphere is also simulated by CORSIKA. CORSIKA allows for the treatment of Cherenkov photons in bunches to allow for more rapid simulations. However, we kept the Cherenkov photon bunch size at 1 in order to avoid the loss of details in the output. Fluctuations in low energy showers can be quite large and bunching Cherenkov photons together can result in the loss of important information in Cherenkov emission profiles and IACT images.

For simulations of the atmosphere, we selected the widely used U.~S.~Standard model with Linsley's parameterisation implemented in CORSIKA. In this model, the atmosphere is represented by five layers. Out of these, the first 4 layers have an exponential density profile, while the last layer (beyond 100~km above sea level) has a linear density profile.

\subsection{Minimum energy threshold}\label{sec_ecuts}

CORSIKA tracks particles in the simulations down to a certain minimum energy threshold. When the energy of a particle falls under that value, the program treats the energy of the particle as having been deposited in the atmosphere through absorption. The value of this minimum cut-off can be set by the user. We have kept the lowest allowed values by CORSIKA for all types of particles: 0.05~MeV for photons and electrons and 50~MeV for hadrons and muons. 

However, in order to evaluate the impact of these thresholds on the longitudinal profiles of gamma-showers, we fixed the photon and electron threshold levels at different values: 50, 5, 3, 0.5, 0.2 and 0.05~MeV. The same threshold is used for both electrons and photons in each case. Since hadrons and muons are only produced in minor processes, the thresholds for them were kept fixed at 50~MeV.

\begin{figure}[h!]
\begin{center}
\includegraphics[width=0.32\textwidth,keepaspectratio]{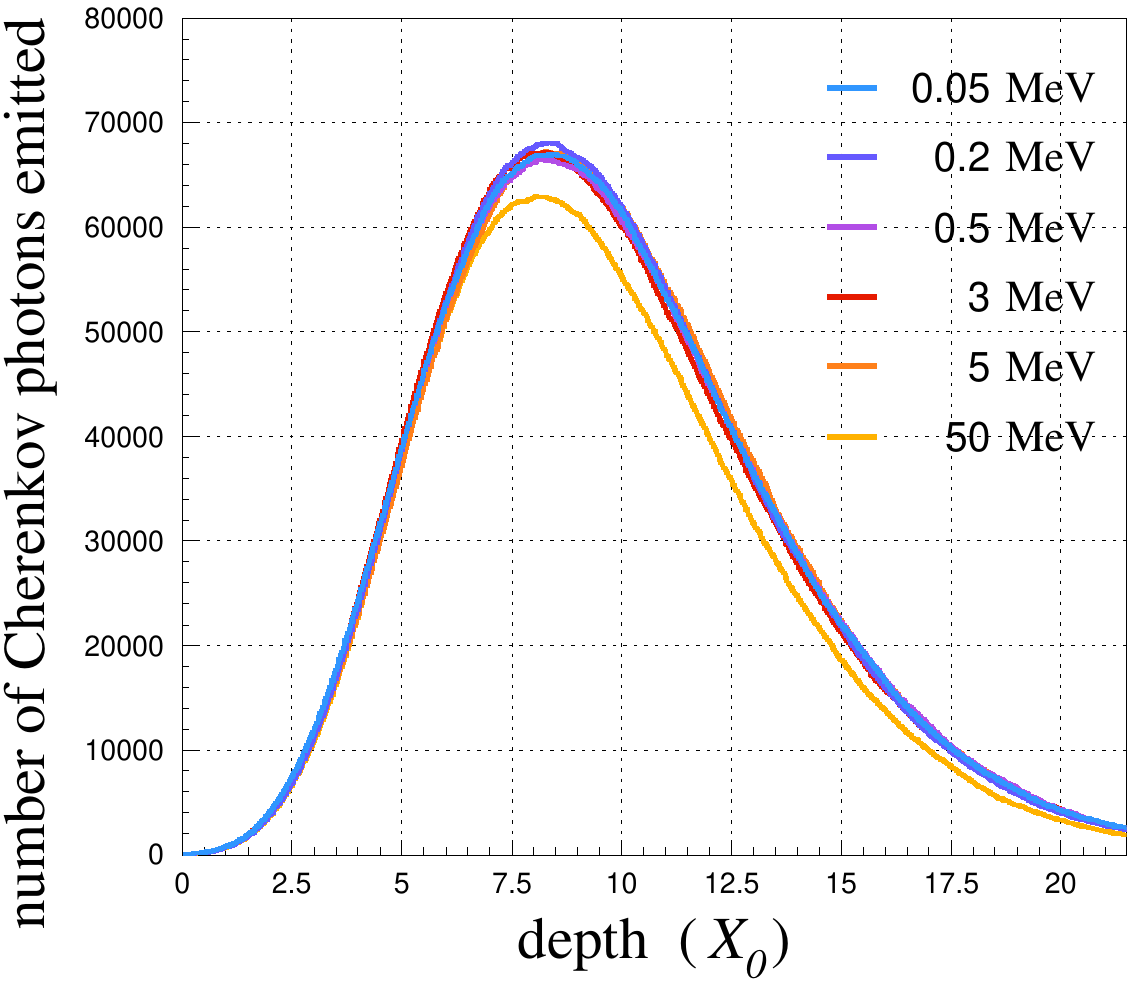}
\includegraphics[width=0.32\textwidth,keepaspectratio]{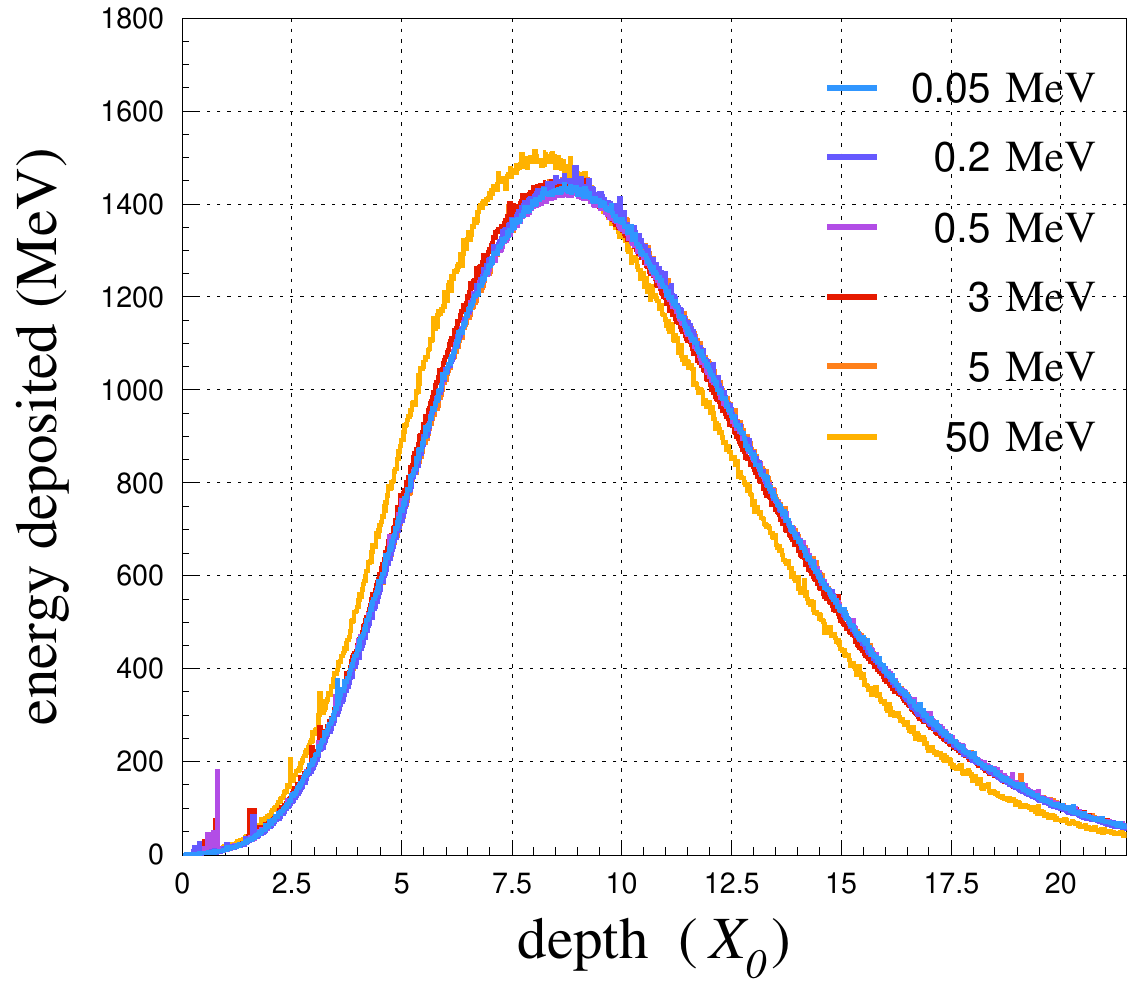}
\includegraphics[width=0.32\textwidth,keepaspectratio]{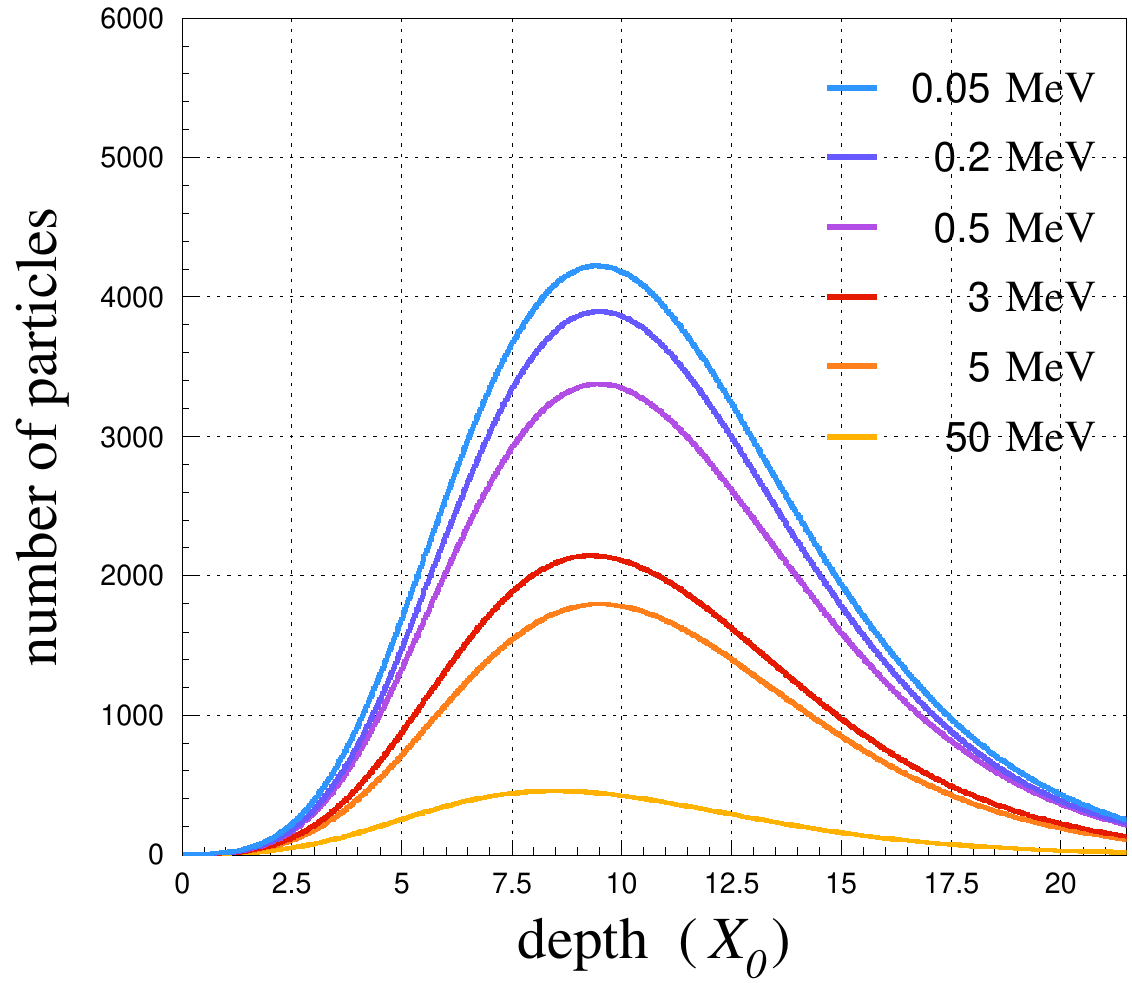}
\caption{\label{fig_ecuts_longi} Average longitudinal profiles for 500~GeV showers as a function of depth (expressed in radiation lengths) obtained for various minimum energy thresholds for electrons and photons used in the CORSIKA simulations are compared. The left plot shows the average Cherenkov emission profiles while the centre plot gives the average profiles of the energy deposited. The right plot shows the average longitudinal profiles in terms of the total number of particles in the shower i.~e. e$^+$,e$^-$ and $\gamma$. The average is carried out over 600 simulated $\gamma$-ray showers at 500~GeV.}
\end{center}
\end{figure}

Figure \ref{fig_ecuts_longi} shows the average longitudinal Cherenkov emission profile obtained for these cuts for 500~GeV showers (top plot). Only the highest of the thresholds tested i.e. 50~MeV has a significant impact on the average profile. All other thresholds yield the same average profile. This can be understood as an effect of the Cherenkov emission threshold in energy. The energy threshold for Cherenkov photon emission is given by equation \ref{eq_cher_ethr} and is lowest at sea level since the refractive index of the atmosphere is highest ($\eta=1.0000293$). At this altitude, electrons need to have an energy of at least 21.1~MeV in order to emit Cherenkov light. At higher altitudes, the threshold will be even higher. For instance, the threshold for Cherenkov emission from electrons is around 50~MeV at an altitude of 15~km above sea level\footnote{For typical showers observed by IACT, significant Cherenkov emission takes place between 5-15~km above sea level.}. Therefore minimum energy cut-off values in the keV range and up to 5~MeV do not have any impact on the Cherenkov photon longitudinal profiles.

For later comparisons, we have also studied the impact of these cuts on the longitudinal profiles for energy deposition in the atmosphere and the number of particles in the shower (electrons and photons).

Figure \ref{fig_ecuts_longi} (centre) shows the average longitudinal energy deposition profiles for the 500~GeV showers in the presence of various minimum energy threshold cuts. Similar to the Cherenkov emission profiles, the energy deposition profiles are only affected at the highest threshold of  50~MeV. Since we have used a threshold of 0.05~MeV in our simulations, the results of the energy deposition profile fits are not affected by this.

The average longitudinal profiles in terms of number of particles (bottom plot in figure \ref{fig_ecuts_longi}) show the most sensitivity to the minimum energy threshold applied. In particular, the value of the cut does not only affect the normalisation of the profiles, it also has an impact on the depth at which the maximum of shower development occurs.  The lower the value of the threshold cut, the greater is the depth of the shower maximum in the atmosphere. Since the current study involves obtaining a parameterisation for the Cherenkov emission longitudinal profiles, this does not affect our results. However, this effect needs to be kept in mind while reading section \ref{subsubsec_tmax} where we have compared the parameterisation of the depth of shower maximum obtained for Cherenkov emission longitudinal profiles with those for particle profiles.

\begin{figure*}[t!]
\begin{center}
\graphicspath{{plots/}}
\includegraphics[width=0.56\textwidth,keepaspectratio]{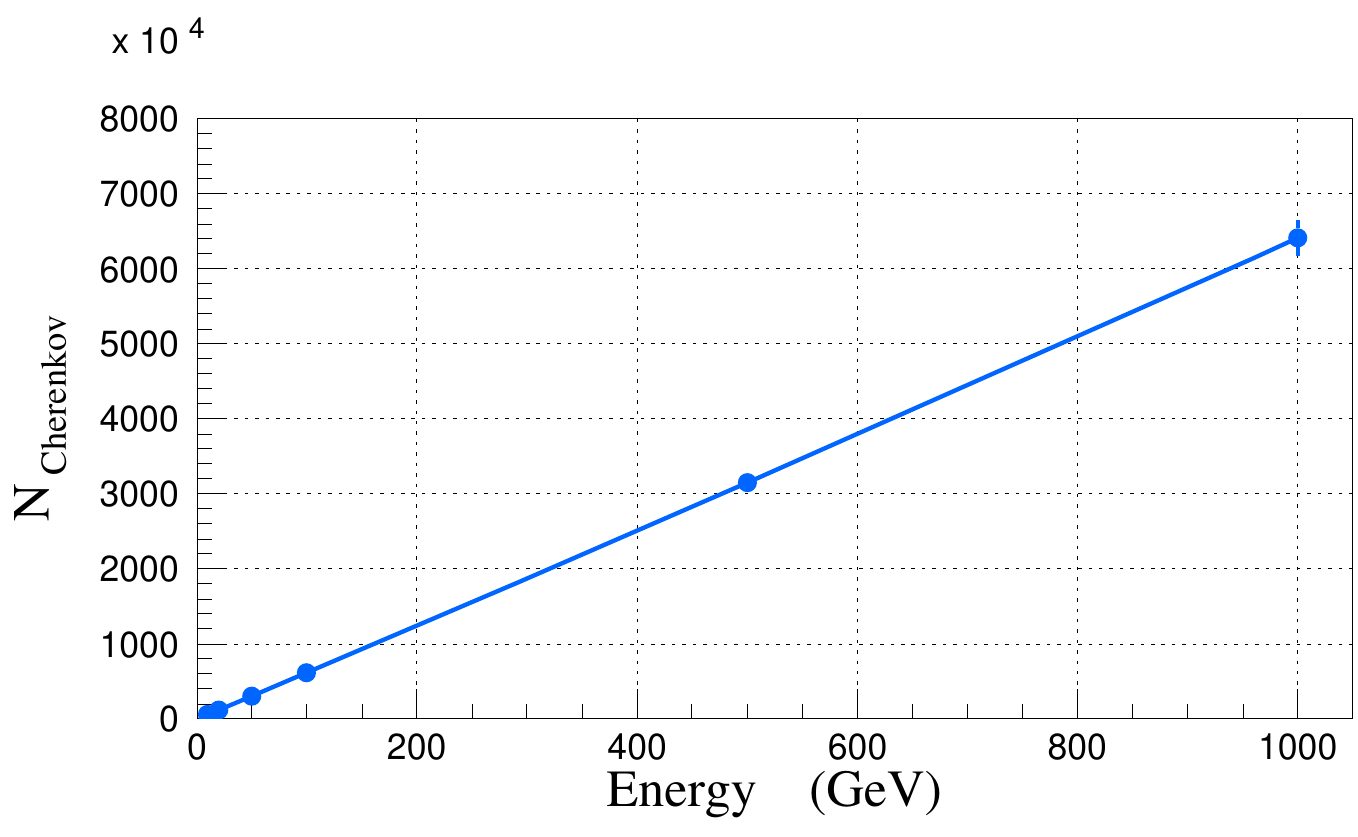}
\hfill
\includegraphics[width=0.38\textwidth,keepaspectratio]{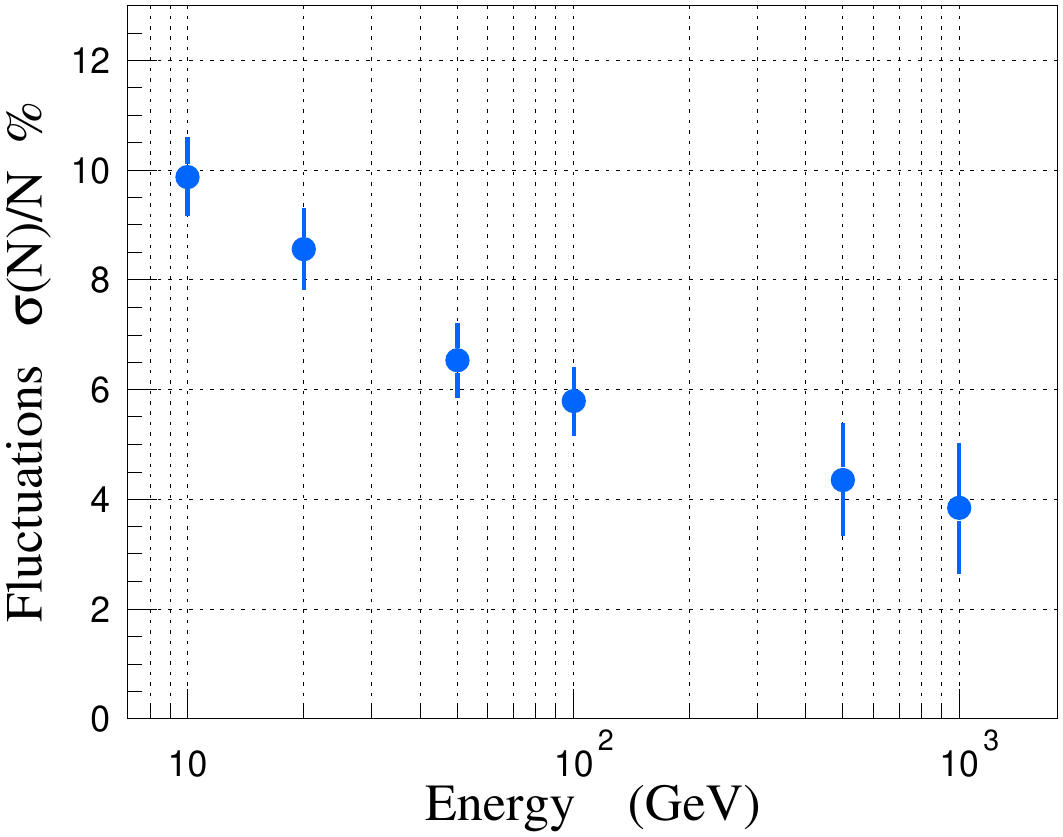}
\caption{\label{fig_photprod} Left: Average number of Cherenkov photons emitted by simulated gamma-ray showers as a function of primary energy. Right: Intrinsic fluctuations of the number of Cherenkov photons emitted in a shower as a function of the energy.}
\end{center}
\end{figure*}

\subsection{Intrinsic fluctuations in the shower and number of simulated showers}
We carried out simulations at 6~gamma-ray energies: 10, 20, 50, 100, 500 and 1000~GeV. 
The number of particles in a shower depends on the energy of the primary. Consequently, the number of Cherenkov photons emitted by a shower also shows the same dependence on the energy Figure \ref{fig_photprod} (left) shows the linear dependence of the average number of Cherenkov photons emitted by gamma-ray showers on the shower energy.

The intrinsic fluctuations in the shower also vary with the energy with lower energy showers showing more variations in their sizes and profiles. In figure \ref{fig_photprod} (right), we show the intrinsic fluctuations in the showers calculated from their Cherenkov emission as a function of shower energy. As we will see in section \ref{sec_results}, the parameters of the gamma function used to describe the longitudinal Cherenkov emission profiles also show a similar inverse dependence on the energy.

In order to account for these fluctuations, the number of showers simulated at each energy was chosen as a function of the energy of the showers. We simulated 10000, 5000, 1000, 500, 200 and 100 showers at 10, 20, 50, 100, 500 and 1000~GeV, respectively. It was determined by gradually increasing the number of showers at 10~GeV and noticing the effect on the average profile of Cherenkov photon distribution on the ground, that the effect of fluctuations are smoothed out by generating 10000 showers. More details can be found in \cite{sajjad:tel-00408835}.

\section{Parameterisation of the Cherenkov emission profiles}\label{sec_results}

In this section, we present the results for the parameterisation of the longitudinal Cherenkov emission profiles for gamma showers simulated in the atmosphere.

\begin{figure*}[htbp]
\begin{center}
\includegraphics[width=0.495\textwidth,keepaspectratio]{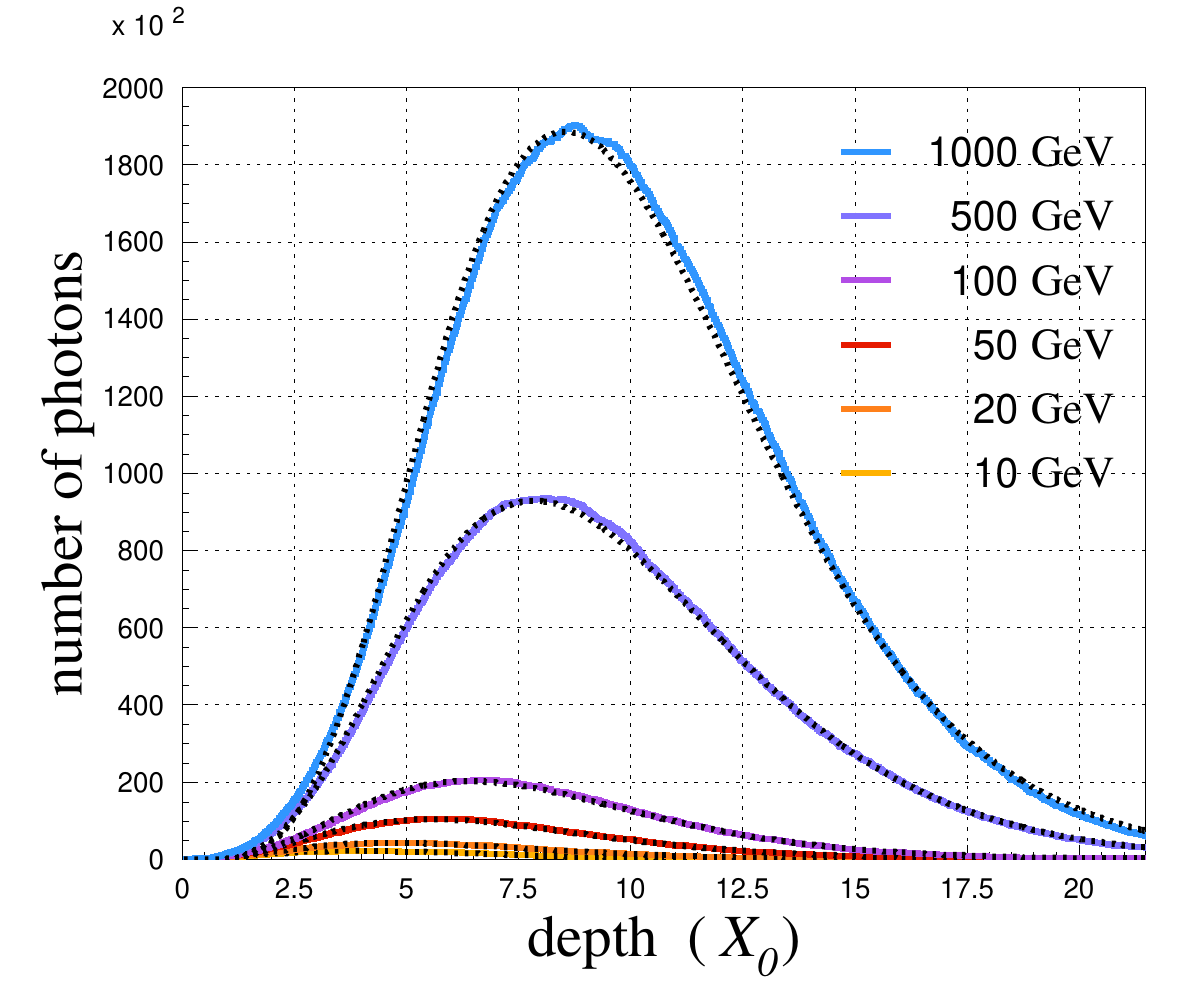}
\hfill
\includegraphics[width=0.495\textwidth,keepaspectratio]{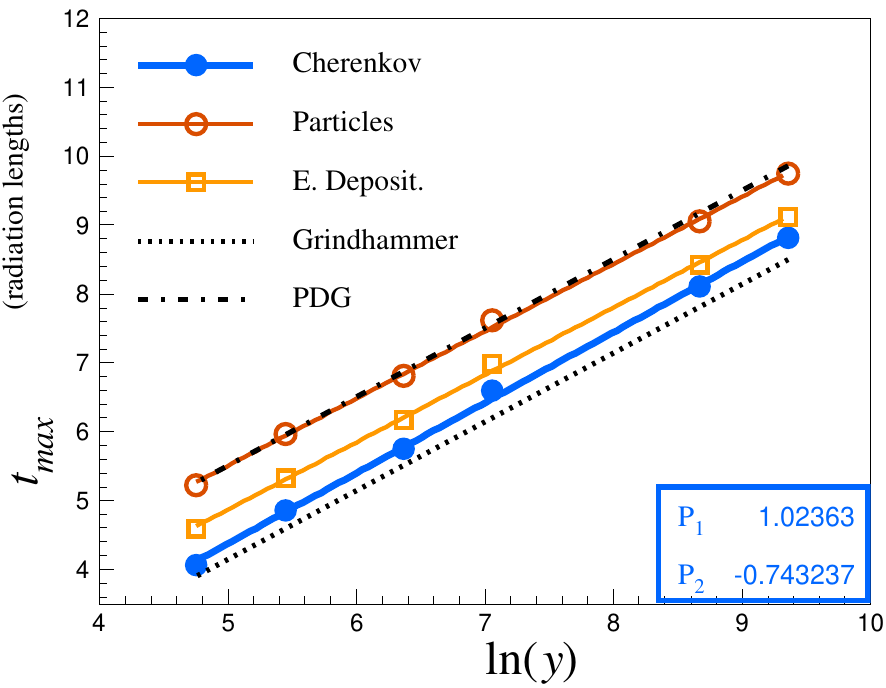}
\caption{\label{fig_fitlongi_cherenk} Left: Average Cherenkov emission longitudinal profiles for gamma showers simulated at various energies (coloured lines). The profiles are fitted with the function $f(t)=C (\beta t)^{\alpha-1}e^{-\beta t}$ (dotted lines). Right: Depth $t_{max}$ of the maximum of average Cherenkov emission profiles (solid points) as a function of the energy of the shower. The depth and energy are expressed in units of radiation length and critical energy, respectively. The parameterisation for $t_{max}$ is obtained by fitting the points with the line f(x)=P$_1$x+P$_2$. The results obtained for the longitudinal profiles in terms of number of particles (electrons and photons) and energy deposited from the same simulations are also shown for comparison. The parameterisations obtained by \cite{grindhammer2000parameterized} and \cite{PhysRevD.98.030001} are also shown.}
\end{center}
\end{figure*}

\subsection{Average profiles}
We first present parameterisations for the average longitudinal Cherenkov emission profiles. Figure \ref{fig_fitlongi_cherenk} (left) shows the average simulated Cherenkov emission profiles for various energies. The profiles are fitted with a function of the form $f(t)=C (\beta t)^{\alpha-1}e^{-\beta t}$ in order to obtain their parameterisation for the gamma function. The fits are also shown in the same figure through the black dotted lines. As discussed in section \ref{subsec_gamma}, the depth is expressed in units of radiation length $X_0$ and the energy is in units of critical energy such that $y=E/E_c$.

\subsubsection{Depth of maximum $t_{max}$} \label{subsubsec_tmax}

The values of the parameters $\alpha$ and $\beta$ obtained from these fits are used to calculate the depth $t_{max}$ of the maximum of longitudinal Cherenkov emission profiles using equation \ref{eq_tmax}. Figure \ref{fig_fitlongi_cherenk} (right) shows the linear dependence of $t_{max}$ on the logarithm of the energy ln($y$) (solid markers). The following parameterisation is obtained by fitting the points with a line.

\begin{equation} \label{eqn_tmax_cher}
	t_{max}=1.024\:\mathrm{ln} (y) -0.743. 
\end{equation}

For comparison, figure \ref{fig_fitlongi_cherenk} (right) also shows the parameterisations obtained by \cite{grindhammer2000parameterized} and \cite{PhysRevD.98.030001} for the energy deposited in the calorimeter medium. The values of the parameters are given  in table \ref{tab_tmax_parameters} for a linear dependence with the general form $t_{max}=\mathrm{P}_1\,\mathrm{ln} (y) + \mathrm{P}_2$.

\begin{table}[h!]
\caption{\label{tab_tmax_parameters}Parameters for the depth of the maximum}
\begin{center}
\begin{tabular}{lcc}
\hline\hline
\multicolumn{1}{c}{}&\small P$_1$ &\small P$_2$ \\
%\hline
%\small 
\hline
\small Cherenkov &  \small 1.024  & \small -0.743\\
\small Particles &\small 0.977 & \small 0.622\\
\small Energy deposited & \small 0.978 & \small -0.023\\
\small Grindhammer et al. & \small 1. & \small -0.858\\
\small PDG & \small 1.  & \small 0.5\\
\hline
\end{tabular}

\vspace{0.cm}
\tablefoot{Values of P$_1$ and P$_2$ for the parameterisations shown in figure  \ref{fig_fitlongi_cherenk} (right). The depth of the maximum for all average longitudinal profiles shows a dependence of the form: $t_{max}=\mathrm{P}_1\,\mathrm{ln} (y) + \mathrm{P}_2$.}
\end{center}
\end{table}

The depth of the Cherenkov emission profile peak shows a closer agreement with \cite{grindhammer2000parameterized} than with \cite{PhysRevD.98.030001}. However, the slopes of the two parameterisations are different. As a result, the deviation between the two becomes larger at higher energies with the average Cherenkov emission profile peak occurring at a deeper depth than the one predicted by \cite{grindhammer2000parameterized} for the same energy. However, the \cite{grindhammer2000parameterized} study obtains parameterisations for energy deposition profiles, while we are studying Cherenkov emission profiles. Moreover, the Grindhammer et al. fits are carried out for homogeneous Copper, Iron, Tungsten, Lead and Uranium media, whereas the current study is focused on electromagnetic showers in air with a variable density profile. In addition, \cite{grindhammer2000parameterized} carry out Monte Carlo simulations through the Geant package, while we have worked with CORSIKA simulations which use the EGS4 package.

The  parameterisation by \cite{PhysRevD.98.030001} is also carried out in several elements between carbon and uranium for energy deposition profiles, through the EGS4 package. This parameterisation shows an energy deposition profile that peaks deeper in the medium compared to the Cherenkov emission profiles. For comparable energies, the depth of the maximum tends to agree more with the depth of the particle longitudinal profile obtained from our simulations in the atmosphere.

Figure \ref{fig_fitlongi_cherenk} also shows the results for average longitudinal profiles for number of particles and energy deposited per unit length (in g/cm$^2$) in the atmosphere. The values of the parameters have been added to table \ref{tab_tmax_parameters}. The peak of the Cherenkov emission profile occurs higher in the atmosphere compared to the energy deposition and particle profiles obtained by the same simulations. The dependence of $t_{max}$ for profiles of electrons only are not shown here, but they were found to closely follow the dependence of the energy deposition profiles.

 Here, a note needs to be made about the longitudinal profiles in terms of number of particles. As described earlier (see section \ref{sec_ecuts}), the minimum energy threshold used in CORSIKA can have an impact on the  profiles in terms of number of particles. The results for these particle longitudinal profiles are therefore indicative of the trend, while the exact values may vary with the application of different minimum energy thresholds. The longitudinal profiles in terms of energy deposited are not affected by this problem.

\begin{figure*}[t!]
\includegraphics[width=0.42\textwidth,keepaspectratio]{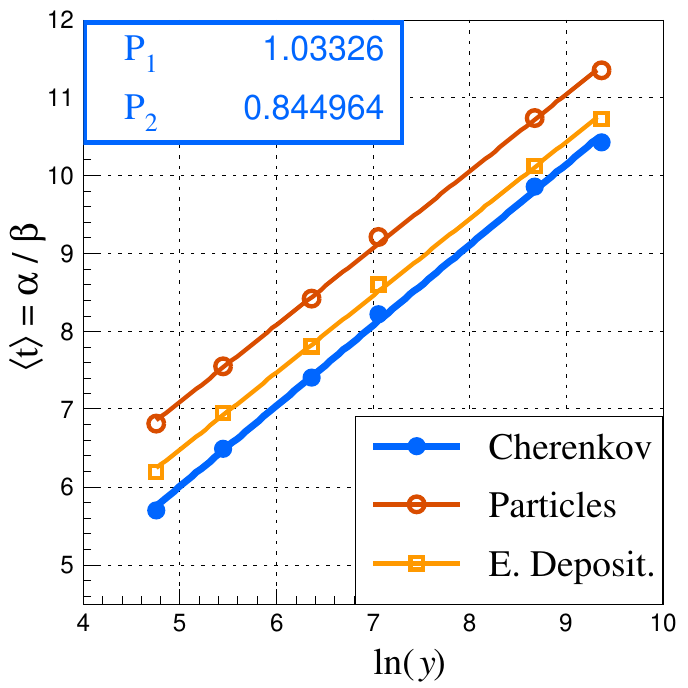}
\hfill
\includegraphics[width=0.42\textwidth,keepaspectratio]{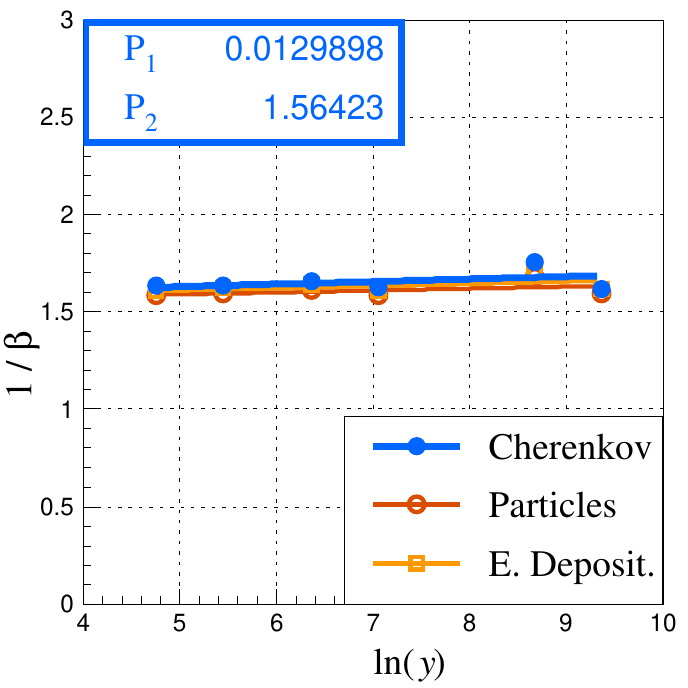}
\caption{\label{fig_fitlongi_cher_parameters} Average value of $\langle t\rangle = \alpha/\beta$ (left) and $1/\beta$ (right) for the longitudinal Cherenkov emission profiles as a function of the shower energy expressed in units of critical energy ($y=E/E_c$) is shown through solid markers. These points are obtained by fitting the profiles in figure \ref{fig_fitlongi_cherenk} with the gamma function and are fitted with a line of the form f(x)=P$_1$x+P$_2$. For comparison, the dependence of average values of $\alpha/\beta$ and $1/\beta$ obtained for particle and energy deposition longitudinal profiles are also shown in the respective plots.}
\end{figure*}

\subsubsection{Mean depth of the profile $\langle t\rangle = \alpha/\beta$}
The values of $\alpha$ and $\beta$ obtained from the  fits of the average longitudinal Cherenkov emission profiles are also used to determine the mean depth of the profile $\langle t\rangle  = \alpha/\beta$. Figure \ref{fig_fitlongi_cher_parameters} (right) shows the values of $\langle t\rangle$ obtained from the fits as a function of the energy of the shower (solid points). This parameter also shows a linear dependence on the logarithm of the energy. The following parameterisation is obtained by fitting the points in the plot.

\begin{equation}\label{eqn_tmean_cher}
\langle t\rangle = \frac{\alpha}{\beta}=1.03\,\mathrm{ln}(y)+0.84
\end{equation}

The comparisons with the average longitudinal particle (hollow circles) and energy deposition (hollow squares) profiles obtained from the same simulations are also shown on the plot. They are discussed below along with the dependence of $1/\beta$ on the energy. 

\subsubsection{The value of $1/\beta$}

Since $t_{max}=\langle t\rangle -\frac{1}{\beta}$, the parameter $1/\beta$ is also expected to have a linear dependence on the logarithm of the energy. Figure \ref{fig_fitlongi_cher_parameters} (right) shows the values of $1/\beta$ obtained from the fits of the average longitudinal Cherenkov emission profiles of showers as a function of the energy (shown through solid markers). The points are fitted with a line and the following parameterisation is obtained.

\begin{equation}\label{eqn_one_o_beta_cher}
	\frac{1}{\beta}=0.01\,\mathrm{ln}(y)+1.58.
\end{equation}

In comparison with the parameters $t_{max}$ and  $\langle t\rangle= \alpha/\beta$ which show a strong dependence on ln$(y)$, the parameter  $1/\beta$ is almost independent of the energy. This results from the fact that the slopes of the parameterisations for $t_{max}$ and  $\langle t\rangle$ have very similar values. This implies that the decaying part of the shower governed by $\beta$ has very little dependence on the energy. Whereas the initial rising part of the shower governed by $\alpha/\beta$ strongly depends on the energy.

The same figure also shows the dependence of $1/\beta$ obtained for the average longitudinal particle and energy deposition profiles (shown through hollow circles and squares respectively). The comparison shows that the value of the average $\beta$ parameter and therefore the decaying part of the profile is almost the same for all types of profiles. In contrast, $\alpha/\beta$ (left plot) has different values for average Cherenkov, particle and energy deposition profiles. This implies that the value of $\alpha$, and hence the rising part of the profile, changes with the longitudinal profile type. The variation of the position of the maximum of the longitudinal profile is also due to the variation in this parameter (via equation \ref{eq_tmax}) since  $\beta$ varies very little and hence does not change the value of $t_{max}$.

\subsection{Fluctuations} \label{subsec_fluctuations}
In this section, we describe the approach used for the estimation of fluctuations, the choice of parameters used to evaluate them and the results of the parameterisations.

\subsubsection{Approach}

 The longitudinal profiles of low energy showers (specially 10 and 20~GeV) fluctuate a lot and fitting them with the gamma function does not always result in convergence. This means that it is harder to obtain values of $\alpha$ and $\beta$ by directly fitting each individual profile. The values of $\alpha$ and $\beta$ can instead be estimated through the following method. The longitudinal Cherenkov emission profile for each shower is divided into mass thickness bins of 1 g/cm$^2$. The depth of the bin with the maximum number of Cherenkov photons emitted is taken as $t_{max}$. Similarly, $\langle t\rangle$ is obtained by taking the weighted average of the Cherenkov emission of all bins along the longitudinal profile. Then equations  \ref{eq_tmax} and \ref{eq_tmean} are used to obtain the value of $\alpha$ and $\beta$.

We note that this method makes the values $\alpha$ and $\beta$ for individual showers sensitive to the fluctuations in the shower itself. The values of these parameters for individual showers are therefore not expected to be exactly those obtained from a converging fit. However, when this process is carried out for a large number of showers, the effects are evened out. As discussed earlier, the number of showers simulated at each energy was chosen to be large enough for the effect of individual shower fluctuations to be smoothed out over averaged behaviour. We, therefore, use the distributions for $\alpha$ and $\beta$ obtained from the method just described to obtain an estimation of the fluctuations in the Cherenkov emission profiles. A similar approach was used by \cite{saadi1989performances} to evaluate the parameters $\alpha/\beta$ and $1/\beta$ for individual electromagnetic showers in the ALEPH calorimeter.

It should be noted that at higher energies such as 500~GeV and 1000~GeV where IACT operate, the longitudinal Cherenkov emission profiles are expected to be easier to fit directly with the gamma function. However, since our purpose here is to obtain a parameterisation for the fluctuations, we use the same method to characterise the fluctuations at all energies.

\begin{figure*}[t!]
\begin{center}
\includegraphics[width=0.325\textwidth,keepaspectratio]{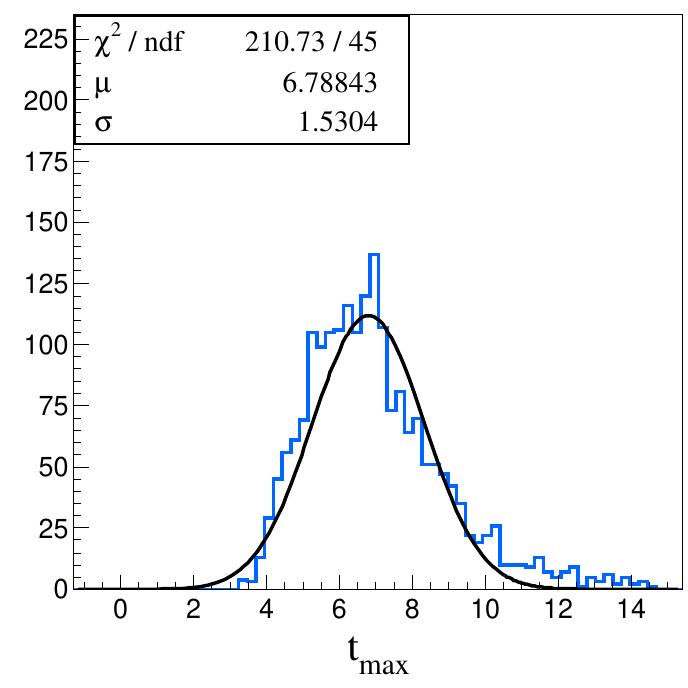}
\includegraphics[width=0.325\textwidth,keepaspectratio]{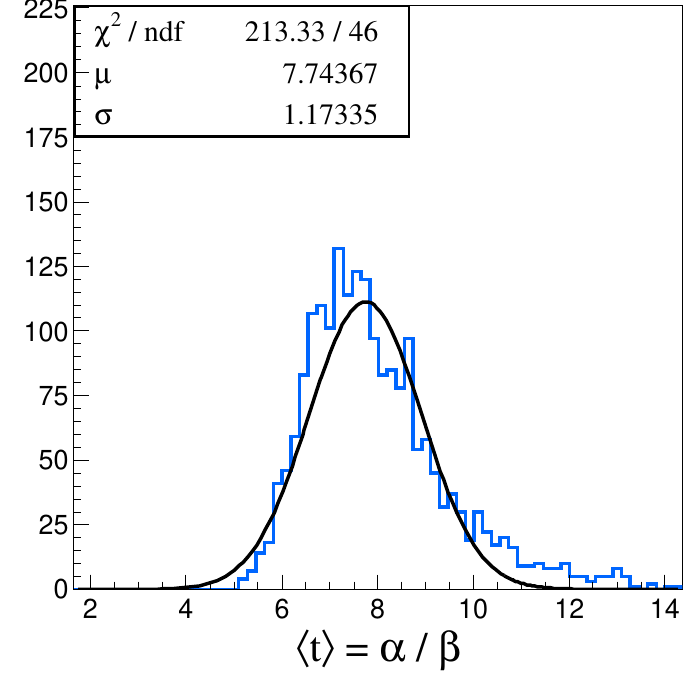}
\includegraphics[width=0.325\textwidth,keepaspectratio]{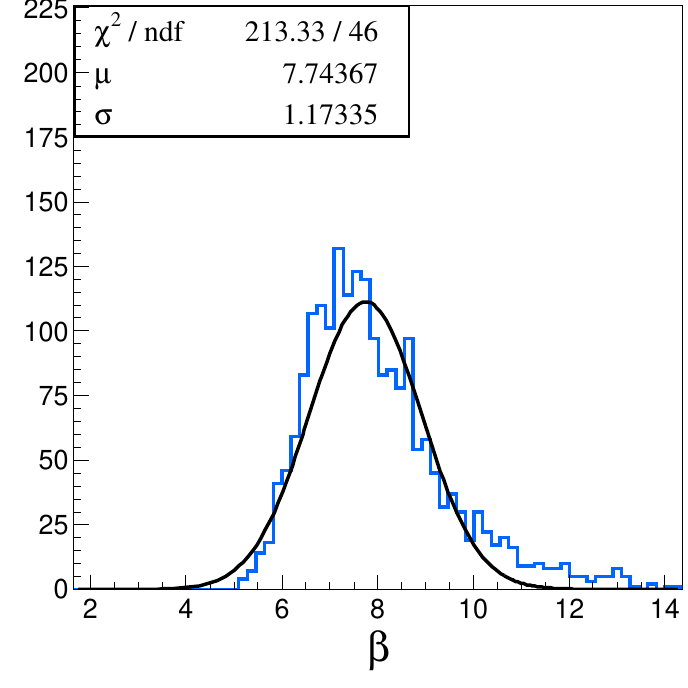}

\vspace{0.3cm}
\includegraphics[width=0.325\textwidth,keepaspectratio]{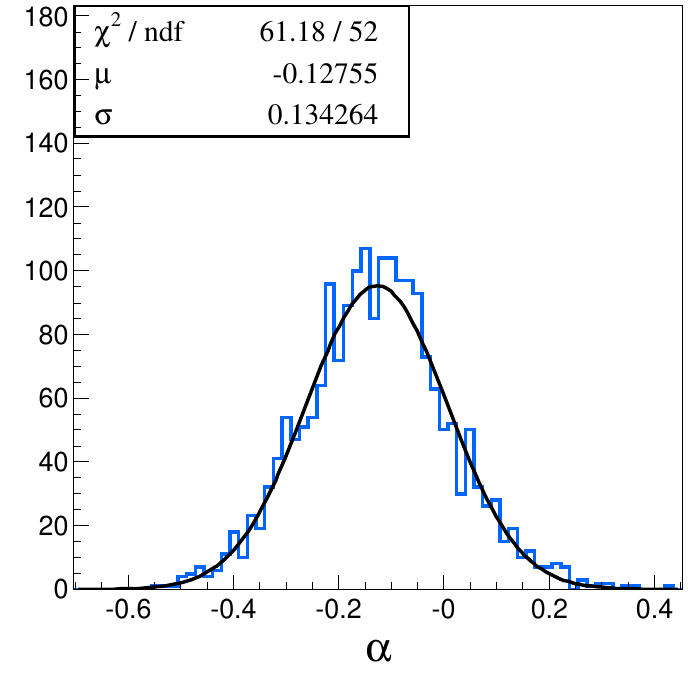}
\includegraphics[width=0.325\textwidth,keepaspectratio]{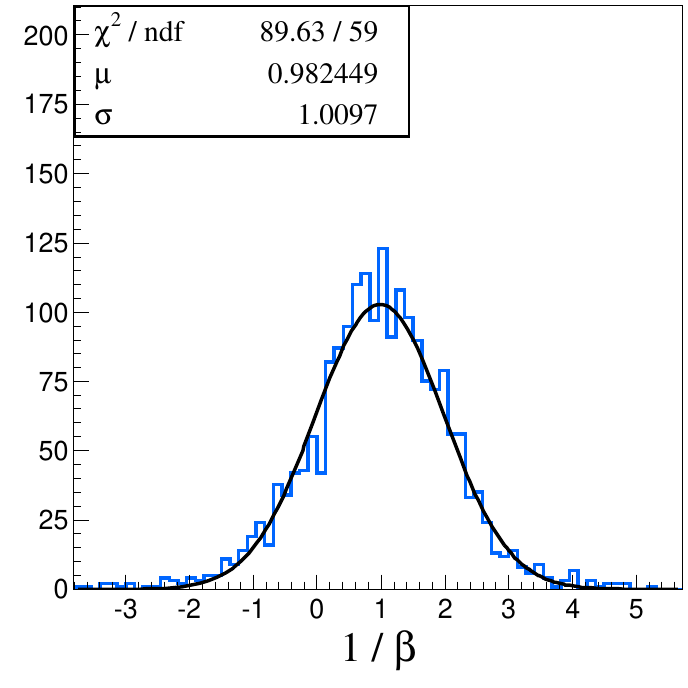}
\includegraphics[width=0.325\textwidth,keepaspectratio]{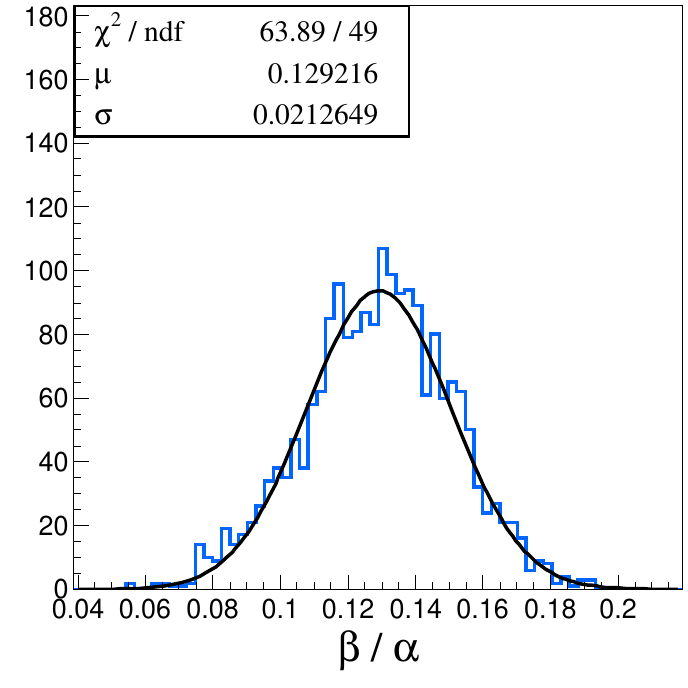}
\end{center}
\begin{minipage}[t]{0.38\textwidth}
\includegraphics[width=0.86\textwidth,keepaspectratio]{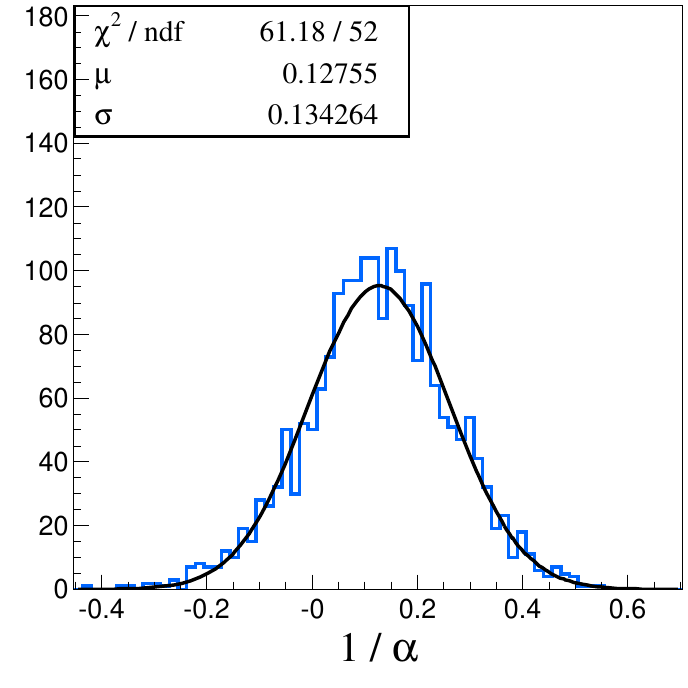}
\end{minipage}
\raisebox{03.65cm}{
\begin{minipage}[t]{0.58\textwidth}
\caption{\label{fig_distributions_symmetrical} Distributions of various parameters obtained from 100~GeV showers (histogram). From left to right, the top gives the distributions of  $t_{max}$,  $\langle t\rangle= \alpha/\beta$ and $\beta$. The centre row shows the distributions of $\alpha$, $1/\beta$, $\beta/\alpha$. While the bottom row gives the distribution for $1/\alpha$. Each distribution is also fitted with a Gaussian function of the form $f(x)=A x^{-0.5(\frac{x-\mu}{\sigma})^2}$ (black line).}
\end{minipage}
}
%}
\end{figure*}

\subsubsection{Choice of parameters: distributions and correlations}
In order to obtain a parameterisation of the fluctuations, we chose to work with parameters that 1) have Gaussian or quasi-Gaussian distributions and 2) are not correlated with each other. The first condition enables us to obtain the values of the mean and standard deviation of the parameter through a Gaussian fit to get an estimate of the fluctuations. The second condition ensures that the parameters can be used independently to fit or simulate shower profiles.

In figure \ref{fig_distributions_symmetrical}, we show the distributions of various parameters obtained from the simulated showers. Only the distributions of 100~GeV showers are shown here, but those at other energies show the same behaviour. The distributions of $t_{max}$,  $\alpha/\beta$ and $\beta$ shown in the top row are all asymmetrical. In contrast, the parameters $\alpha$, $1/\beta$, $\beta/\alpha$ and $1/\alpha$, shown in the centre and bottom rows all have Gaussian or quasi-Gaussian distributions and have the potential to be used to quantify the fluctuations on the profiles.

\begin{figure*}[t!]
\begin{center}
\includegraphics[width=0.32\textwidth,keepaspectratio]{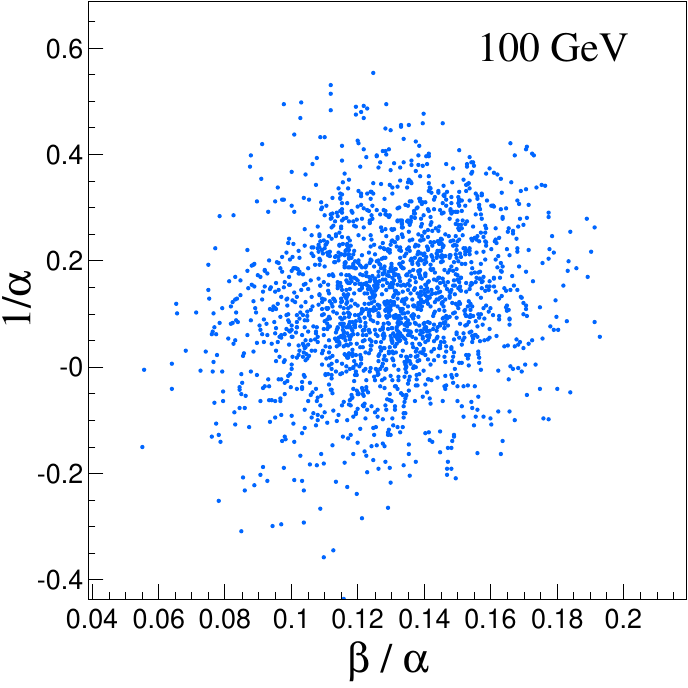}
\includegraphics[width=0.32\textwidth,keepaspectratio]{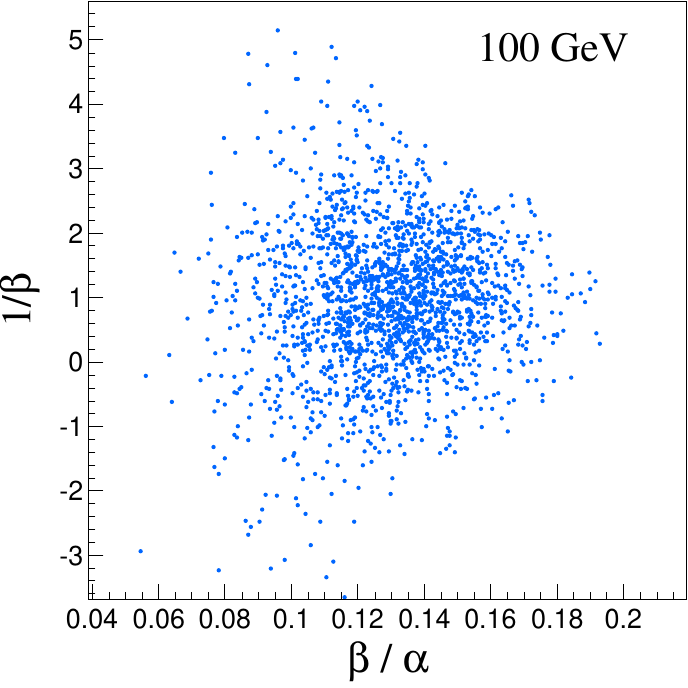}
\includegraphics[width=0.32\textwidth,keepaspectratio]{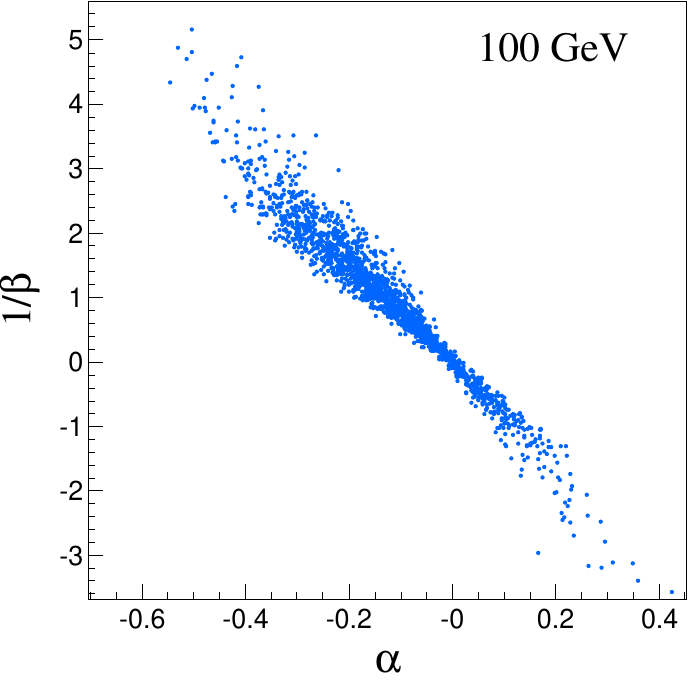}

\vspace{0.5cm}
\includegraphics[width=0.32\textwidth,keepaspectratio]{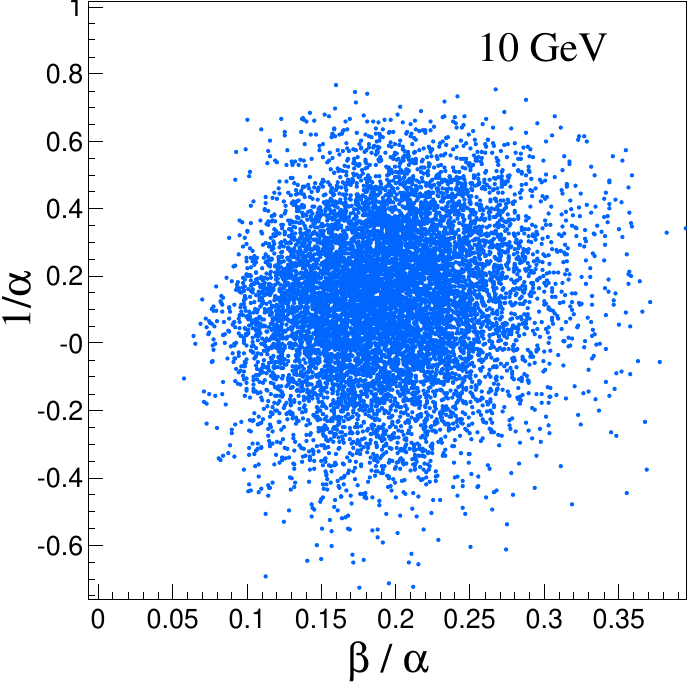}
\includegraphics[width=0.32\textwidth,keepaspectratio]{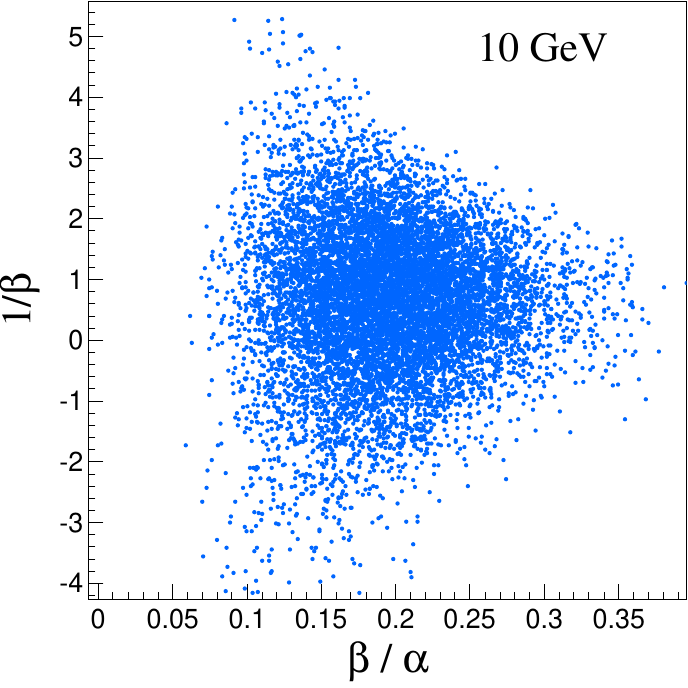}
\includegraphics[width=0.32\textwidth,keepaspectratio]{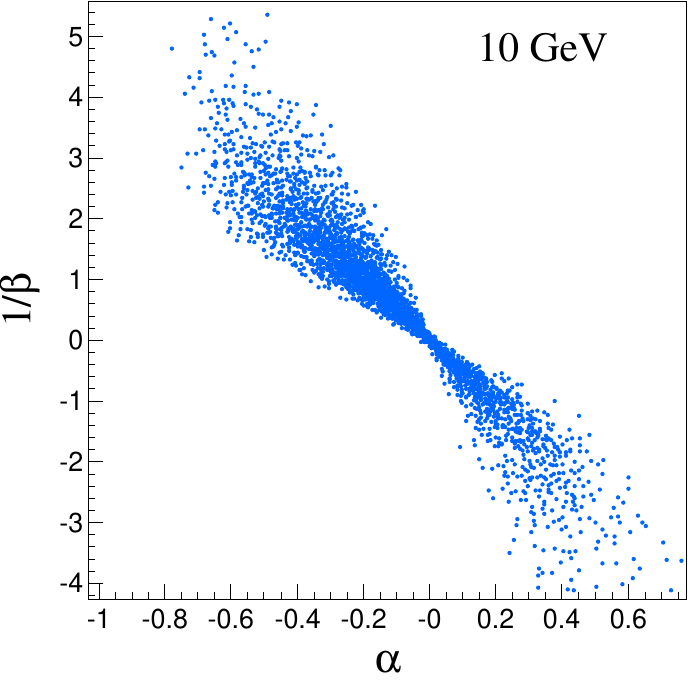}
\caption{\label{fig_correlations} Top row: Correlations between $1/\alpha$ and $\beta/\alpha$ (left),  $1/\beta$ and $\beta/\alpha$ (centre), and $1/\beta$ and $\alpha$ (right) for 100~GeV simulated showers. Bottom row: The same correlations for 10~GeV simulated showers.}
\end{center}
\end{figure*}

We next study the correlations between these four parameters. Figure \ref{fig_correlations} (top row) shows the relationship between the following pairs of parameters for 100~GeV showers:  $1/\alpha$ and $\beta/\alpha$ (left),  $1/\beta$ and $\beta/\alpha$ (centre), and $1/\beta$ and $\alpha$ (right). In the bottom row, we also give the same distributions for 10~GeV showers since the correlations are more evident with a larger number of simulated showers. The distributions from all other energies showed the same trend as these two. These plots show that $1/\alpha$ and $\beta/\alpha$ are the only pair of parameters that are independent from each other. In comparison, $1/\beta$ and $\beta/\alpha$ have some correlation between them. While $1/\beta$ and $\alpha$ are strongly correlated implying that $1/\beta$ and $1/\alpha$ will also be correlated.

Since $\beta/\alpha$ and $1/\beta$ are the only parameters with Gaussian distributions and no correlation with each other, we use them to estimate the fluctuations of the longitudinal profiles in the following section. \cite{saadi1989performances} also chose the same parameters to evaluate fluctuations in the electromagnetic showers for the ALEPH calorimeter. 

The distributions and correlations of these two parameters for all energies are given in \ref{app_fluctuations}  and \ref{app_correlation}, respectively.

\subsubsection{The parameterisation of fluctuations}

\begin{figure*}[t!]
\begin{center}
\includegraphics[width=0.95\textwidth,keepaspectratio]{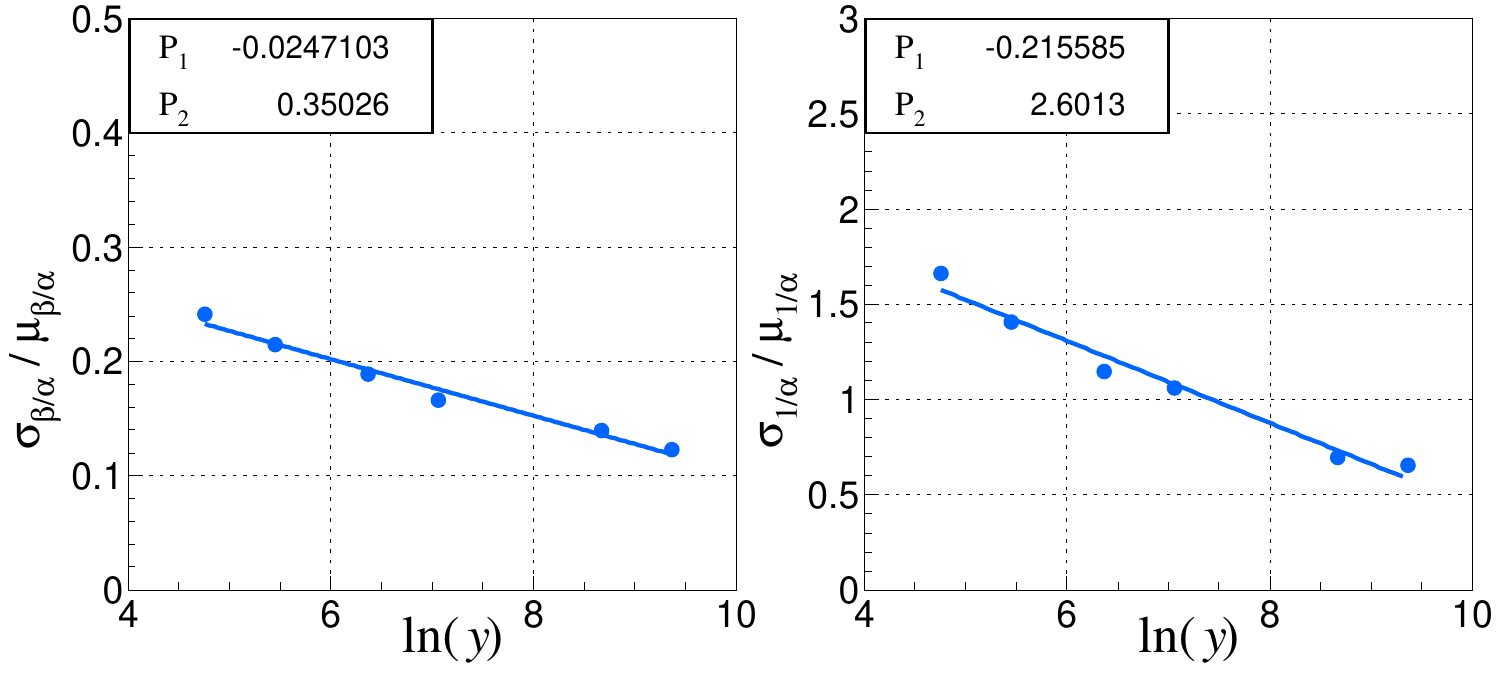}
\caption{\label{fig_fitlongi_cher_sigmav} Dependence of $\sigma_{\beta/\alpha}/\mu_{\beta/\alpha}$ (left)  and $\sigma_{1/\alpha}/\mu_{1/\alpha}$ (right) on the energy of the showers. The points are obtained from the distributions  of $\beta/\alpha$ and  $1/\alpha$ and fitted with the line f(x)=P$_1$x+P$_2$.}

\end{center}
\end{figure*}

The distributions for $\beta/\alpha$ and $1/\alpha$ are fitted with a Gaussian function as shown in figure  \ref{fig_distributions_symmetrical} for 100~GeV showers. The plots for all energies are given in \ref{app_fluctuations}. The Gaussian fits give us the mean $\mu$, standard deviation $\sigma$ of each distribution. For each energy, the fluctuations follow a linear dependence on the energy and are given by the ratio of $\sigma/\mu$. Figure \ref{fig_fitlongi_cher_sigmav} (left and centre) gives the ratio $\sigma_{\beta/\alpha}/\mu_{\beta/\alpha}$ (left) and $\sigma_{1/\alpha}/\mu_{(1/\alpha)}$ (right) as a function of the energy. These graphs are then fitted with a straight line which yields the following parameterisations for the fluctuations:

\begin{eqnarray}
	\frac{\sigma_{\beta/\alpha}}{\mu_{\beta/\alpha}}&=&-0.025\,\mathrm{ln}(y)+0.35,\\
	\frac{\sigma_{1/\alpha}}{\mu_{1/\alpha}}&=&-0.216\,\mathrm{ln}(y)+2.6.\\
\end{eqnarray}

As expected, the fluctuations on the parameters are higher at lower energies due to the fluctuations of the shower development itself.

\section{Summary, conclusions and future directions}\label{sec_conclusions}

In order to  parameterise longitudinal Cherenkov emission profiles for IACT, we have carried out simulations for gamma-ray showers in the atmosphere through CORSIKA in the GeV-TeV range.
In doing so, we have also studied the effect of varying the minimum energy threshold cut-off for electrons and photons in the CORSIKA simulations.  We have found that the average longitudinal Cherenkov and energy deposition profiles are unaffected by keeping the threshold cut-offs below 5~MeV. In contrast, the average longitudinal particle profiles show variations for all minimum energy threshold cut-off values allowed by CORSIKA. This underlines the need for caution while evaluating particle number longitudinal profiles through CORSIKA.

We have also evaluated the intrinsic fluctuations in the gamma-ray showers through the total number of Cherenkov photons emitted by a single shower and shown that these fluctuations increase at lower energies.
For the parameterisation of longitudinal Cherenkov emission profiles, the  number of showers generated at each energy was chosen so as to smooth out the effects of individual shower fluctuations. 

Using the simulations thus obtained, we have obtained parameterisations for the longitudinal Cherenkov emission profiles for gamma-ray showers in the atmosphere, using the gamma-function. The Cherenkov emission profiles are well represented by the gamma function. The depth of the maximum of the average Cherenkov emission profile, $t_{max}$, shows a linear dependence on the logarithm of the energy. In comparison with the parameterisations obtained for the energy deposition profiles of gamma-ray showers in various materials by \cite{grindhammer2000parameterized} and \cite{PhysRevD.98.030001}, our parameterisation for $t_{max}$ gives results that are closer to those obtained by \cite{PhysRevD.98.030001}. There is however a difference in the slope obtained which means that the results tend to deviate from each other at higher energies. In comparison with the particle and energy deposition profiles obtained from the same simulations, the Cherenkov emission profiles have peaks that occur higher in the atmosphere.

The parameter $\alpha/\beta$ which is equal to the mean depth of the shower $\langle t\rangle$ also shows a linear dependence on the logarithm of the energy for average Cherenkov emission profiles. In comparison, the parameter  $1/\beta$ shows a very mild linear dependence on the logarithm of the energy, showing that the decaying part of the longitudinal Cherenkov emission profile depends very little on energy. Comparisons with the same parameters obtained for particle and energy deposition profiles show that $\alpha/\beta$ varies with the type of profile while $1/\beta$ does not.

In order to evaluate the fluctuations on the longitudinal Cherenkov emission profiles, we have worked with the parameters $\beta/\alpha$ and $1/\alpha$. Both parameters have Gaussian distributions and are not correlated with each other. The fluctuations of both parameters show a linear dependence on the logarithm of energy. Similar to shower fluctuations, these parameters also show more fluctuations at lower energies as compared to higher ones.

The lack of correlations and the Gaussian distribution of $\beta/\alpha$ and $1/\alpha$ mean that these parameterisations can be used to generate gamma-ray shower longitudinal Cherenkov emission profiles in the atmosphere for fast simulations. The fast simulations can be carried out for various energies or spectra of energies. Moreover, since the parameterisations are expressed in units of radiation lengths and critical energy with the gamma function, the profiles generated can also be adapted to different seasons, latitudes and other atmospheric conditions provided the atmospheric density profile is known.

In future works, such parameterisations can also be carried out for reconstructed longitudinal profiles for images obtained from IACT. In general, the longitudinal Cherenkov emission profile depends on two parameters of the shower: the energy and the depth of first interaction. Methods for determining the energy of the shower already exist; therefore such parameterisations can potentially be used to better reconstruct the longitudinal profile in the atmosphere and position of the source in the sky. Moreover, since the longitudinal profiles of hadronic showers are more irregular and their development takes place deeper in the atmosphere and over longer distances due to the greater nuclear interaction length, longitudinal profiles can also be used to improve discrimination between hadronic and electromagnetic showers. Future works would then involve using these parameterisations to further develop IACT image analysis techniques or the development of new techniques.

\section{Acknowledgements}

The authors would like to acknowledge the support of the Languedoc-Roussillon region (Bourse CNRS BDI-PED ) for this research project. The authors are also grateful to Georges Vasileiadis and Julien Bolmont for reading the manuscript and making valuable comments.

\section*{References}
\bibliographystyle{bibtex/aa}
\bibliography{article}

\begin{appendix}

\section{Fluctuations}\label{app_fluctuations}

\begin{center}
\begin{minipage}[t]{0.99\textwidth}
\begin{center}
\includegraphics[width=0.25\textwidth,keepaspectratio]{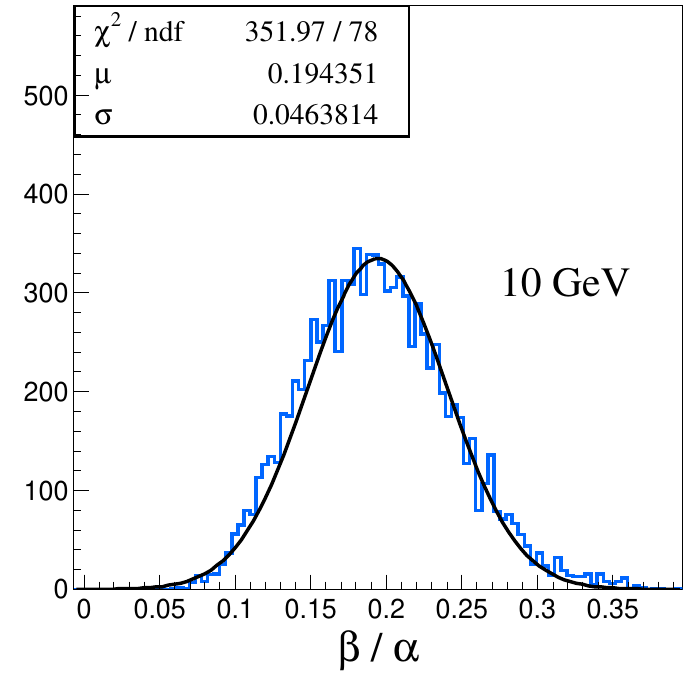}
\includegraphics[width=0.25\textwidth,keepaspectratio]{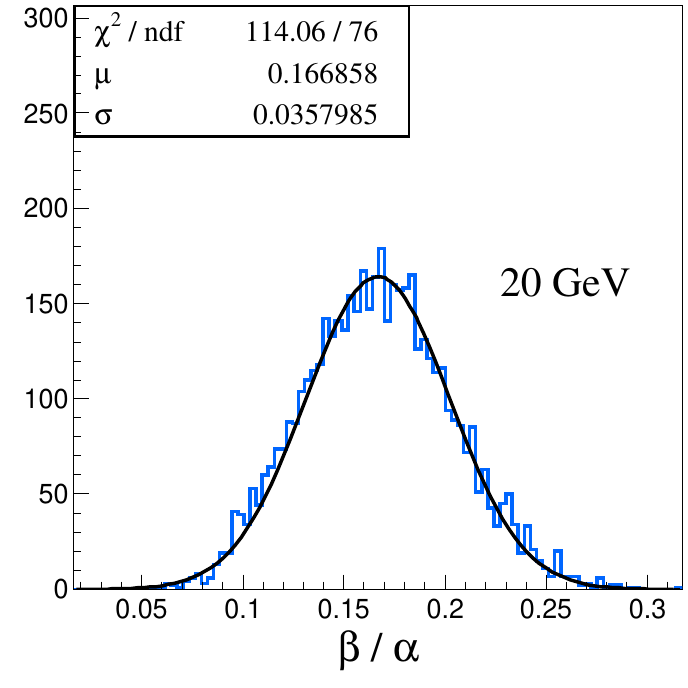}
\includegraphics[width=0.25\textwidth,keepaspectratio]{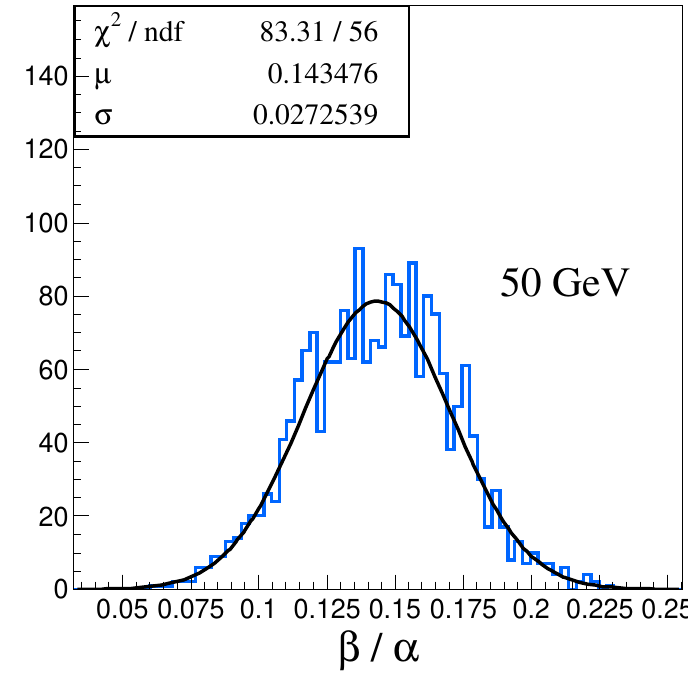}
\end{center}
\end{minipage}
\begin{minipage}[t]{0.99\textwidth}
\begin{center}
\includegraphics[width=0.25\textwidth,keepaspectratio]{fitlongi_cher_parameters_beta_o_alpha_100.pdf}
\includegraphics[width=0.25\textwidth,keepaspectratio]{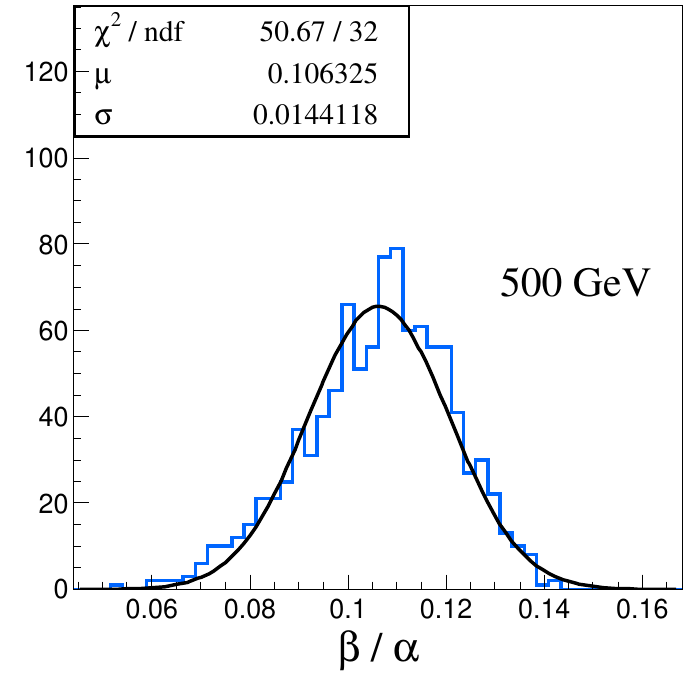}
\includegraphics[width=0.25\textwidth,keepaspectratio]{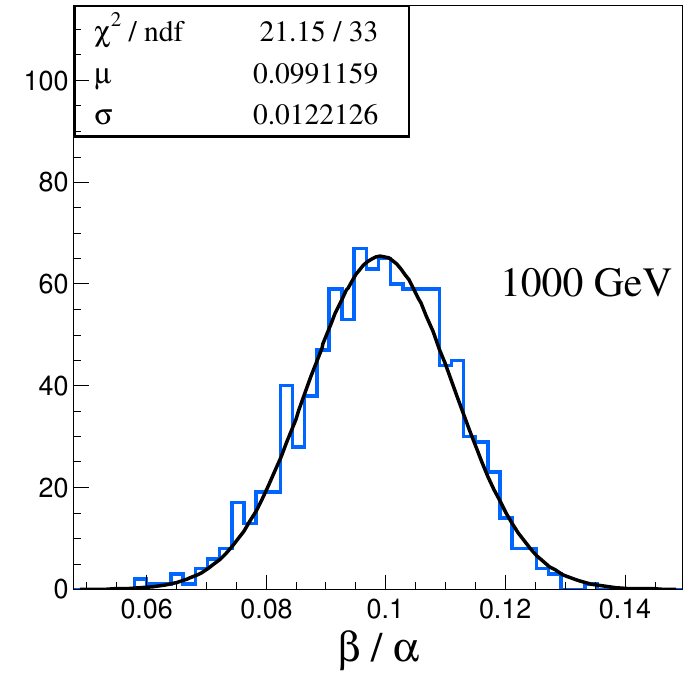}
\end{center}
\captionof{figure}{Distributions of $\beta/\alpha$ for various energies. The distributions are fitted with a Gaussian function of the form $f(x)=A x^{-0.5(\frac{x-\mu}{\sigma})^2}$ (black line).}\label{fig_fitlongi_cher_gaussian_beta_o_alpha}
\end{minipage}
\end{center}

\begin{center}
\begin{minipage}[t]{0.99\textwidth}
\begin{center}
\includegraphics[width=0.25\textwidth,keepaspectratio]{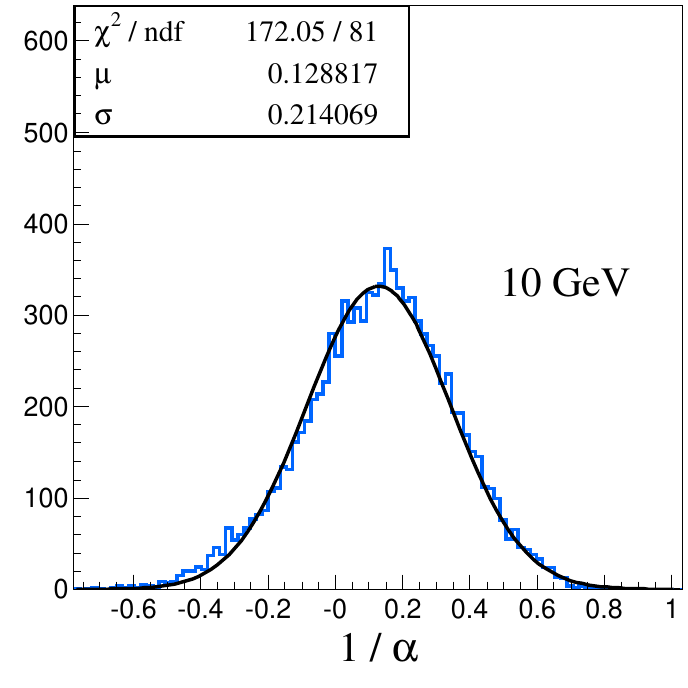}
\includegraphics[width=0.25\textwidth,keepaspectratio]{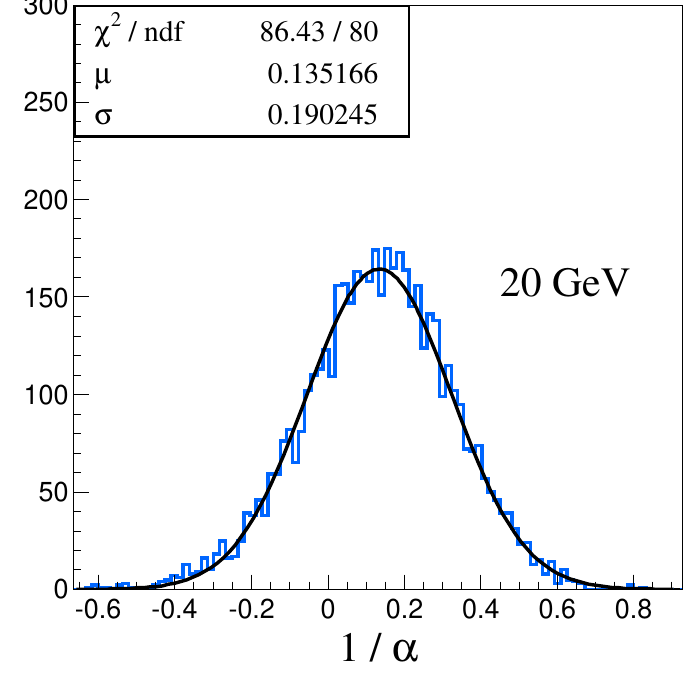}
\includegraphics[width=0.25\textwidth,keepaspectratio]{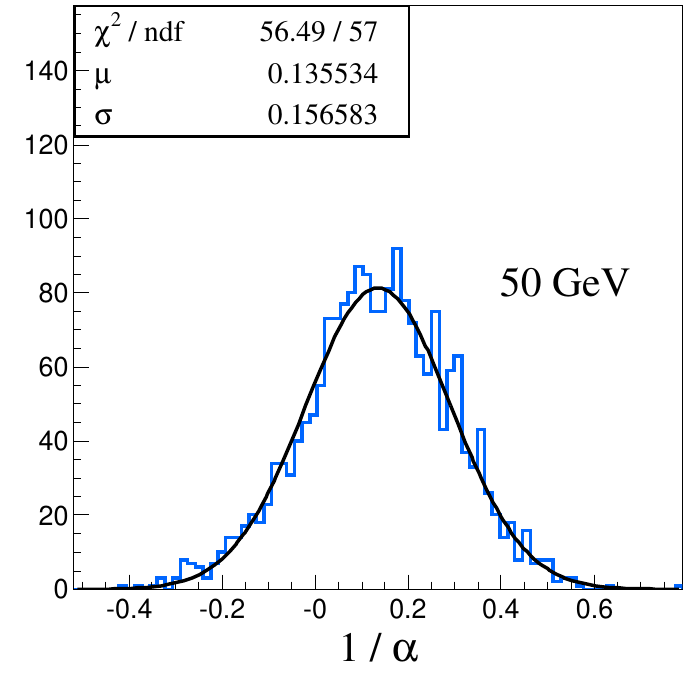}
\end{center}
\end{minipage}
\begin{minipage}[t]{0.99\textwidth}
\begin{center}
\includegraphics[width=0.25\textwidth,keepaspectratio]{fitlongi_cher_parameters_one_o_alpha_100.pdf}
\includegraphics[width=0.25\textwidth,keepaspectratio]{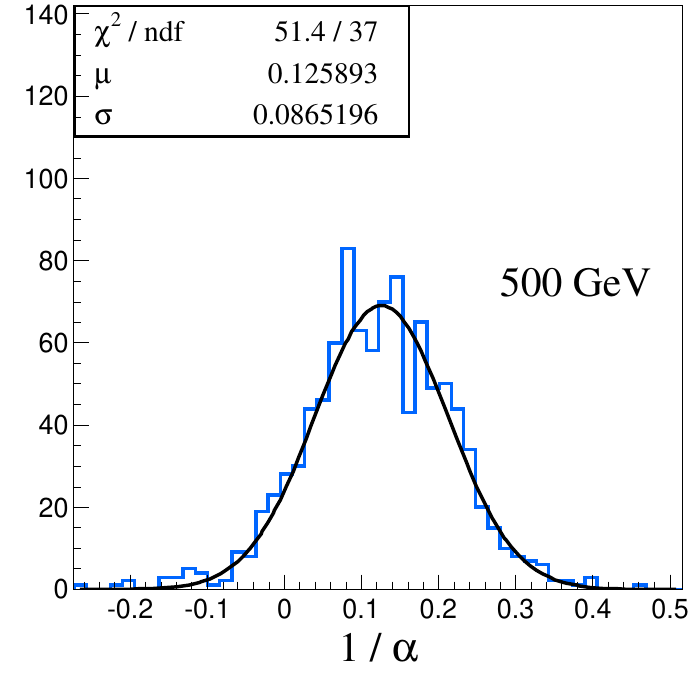}
\includegraphics[width=0.25\textwidth,keepaspectratio]{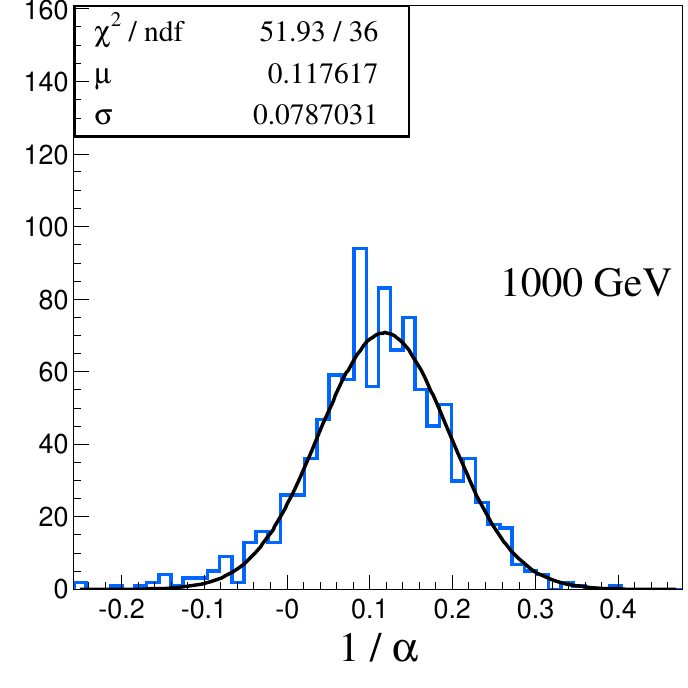}
\end{center}
\captionof{figure}{Distributions of $1/\alpha$ for various energies. The distributions are fitted with a Gaussian function of the form $f(x)=A x^{-0.5(\frac{x-\mu}{\sigma})^2}$ (black line).}\label{fig_fitlongi_cher_gaussian_one_o_alpha}
\end{minipage}
\end{center}

\clearpage

\section{Correlation between parameters}\label{app_correlation}

\begin{center}
\begin{minipage}[t]{0.99\textwidth}
\begin{center}
\includegraphics[width=0.25\textwidth,keepaspectratio]{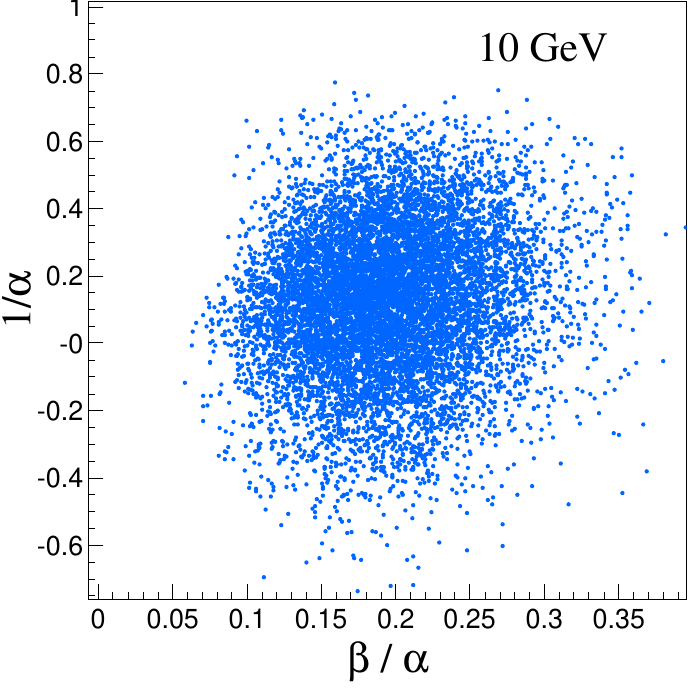}
\includegraphics[width=0.25\textwidth,keepaspectratio]{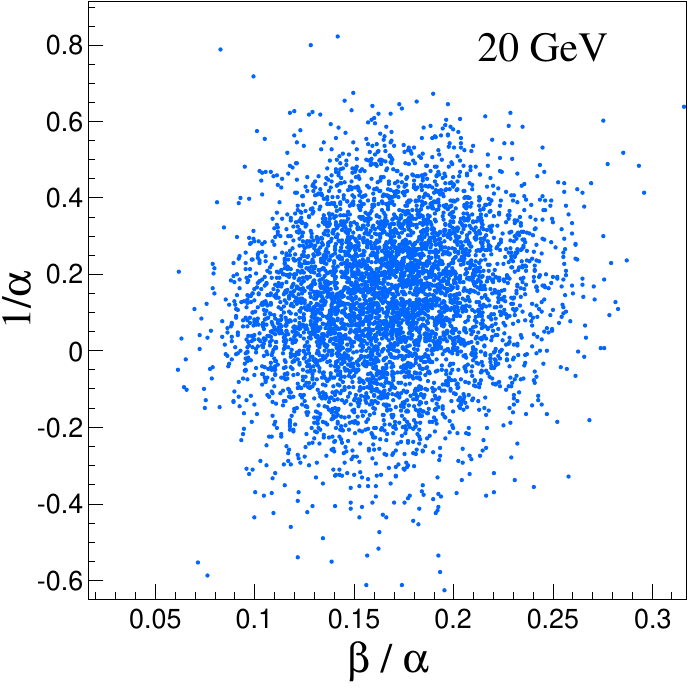}
\includegraphics[width=0.25\textwidth,keepaspectratio]{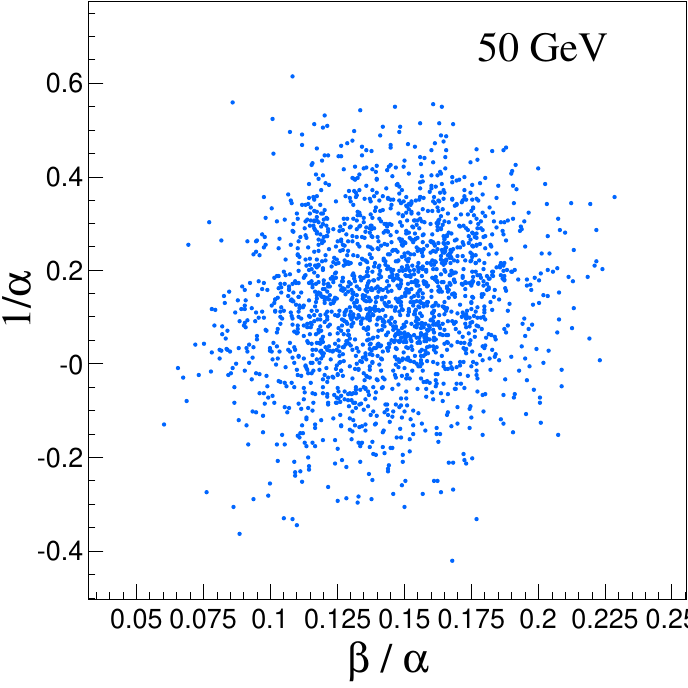}
\includegraphics[width=0.25\textwidth,keepaspectratio]{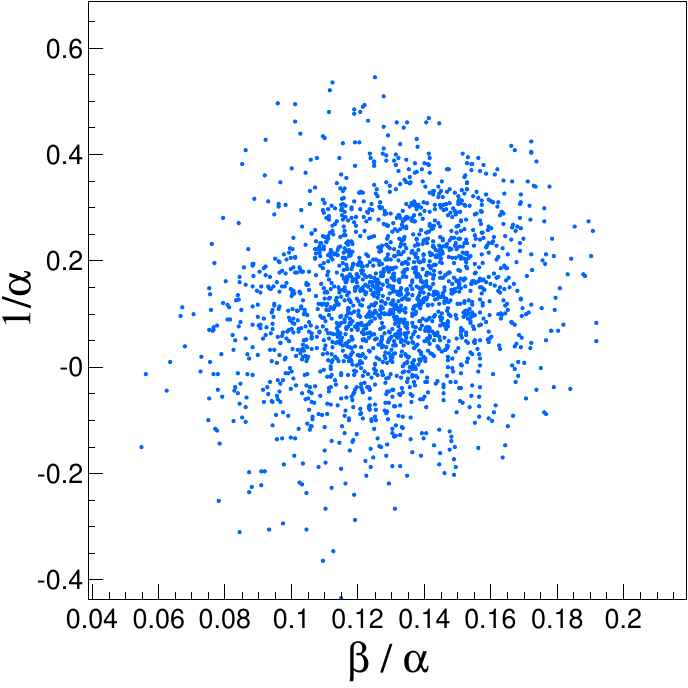}
\includegraphics[width=0.25\textwidth,keepaspectratio]{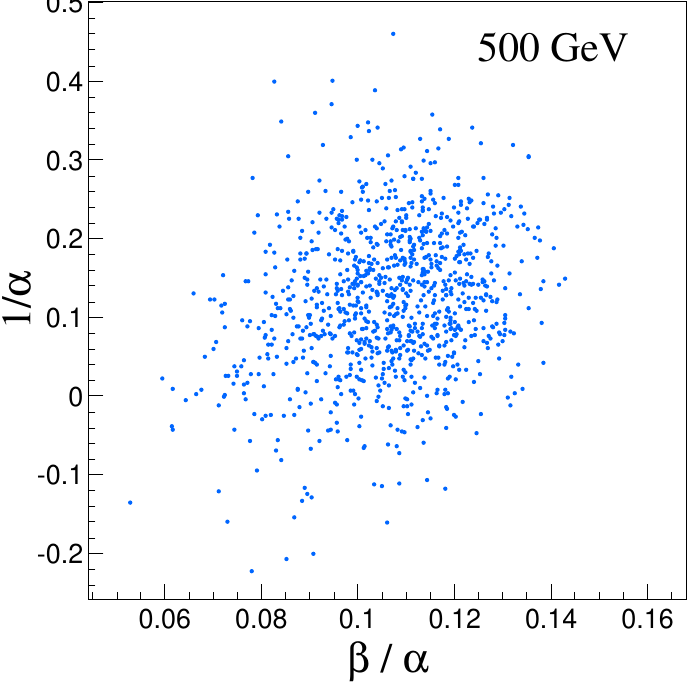}
\includegraphics[width=0.25\textwidth,keepaspectratio]{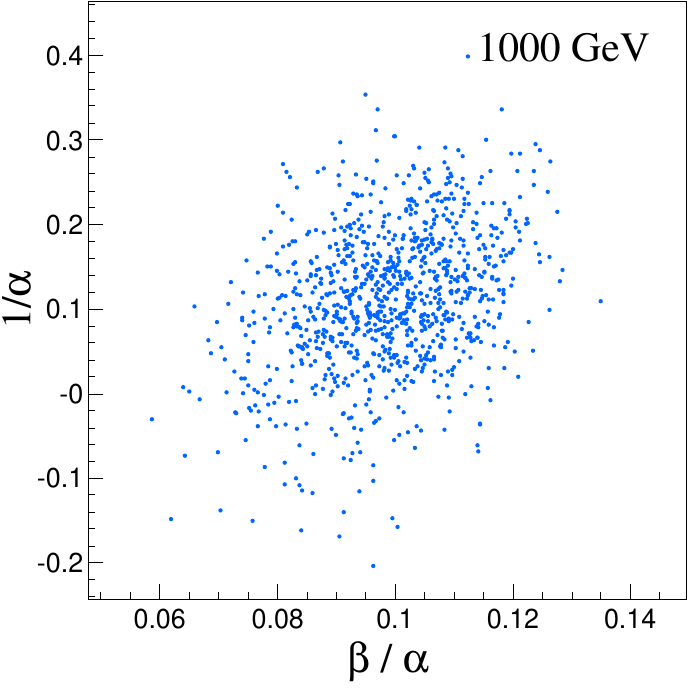}
\end{center}
\captionof{figure}{Scatter plots of $1/\alpha$ versus $\beta/\alpha$ for various energies.}\label{fig_fitlongi_cher_gaussian_correlation}
\end{minipage}
\end{center}

\end{appendix}

\end{document}